\documentclass{article}[12pt]
\pdfoutput=1

\usepackage[utf8]{inputenc}
\usepackage[T1]{fontenc}

\usepackage[margin=2.5cm]{geometry}
\usepackage{parskip}
\usepackage{setspace}

\usepackage{amsmath}
\usepackage{amssymb}
\usepackage{mathtools}
\usepackage{slashed}

\usepackage{hyperref}

\usepackage{graphicx}
\usepackage{color}

\usepackage{listings}

\definecolor{darkRed}{RGB}{160,0,0}
\definecolor{darkGreen}{RGB}{0,160,0}
\definecolor{darkMagenta}{RGB}{120,60,0}
\definecolor{darkCyan}{RGB}{0,100,100}

\hypersetup{
	pdftitle={SpinorsExtras},
	pdfauthor={Jakub Kuczmarski},
	pdfsubject={%
		Mathematica implementation of massive spinor-helicity formalism%
	},
	pdfcreator={Jakub Kuczmarski},
	pdfproducer={Jakub Kuczmarski},
	pdfkeywords=%
		{spinor helicity} {massive spinor} {on-shell recursion} {Mathematica},
	colorlinks=true,		
	linkcolor=darkRed,		
	citecolor=darkGreen,	
	filecolor=darkMagenta,	
	urlcolor=darkCyan		
}

\DeclareMathOperator{\sgn}{sgn}

\newcounter{tutorialStep}
\newcommand{\tutorialStep}{%
	\addtocounter{tutorialStep}{1} \noindent \textbf{\arabic{tutorialStep})~}%
}

\begin{document}

\begin{titlepage}

\begin{flushright}
IFT/14/03
\end{flushright}

\vspace{4em}

\begin{center}
\begin{doublespace}
\textbf{\Large
	SpinorsExtras -
	Mathematica implementation of massive spinor-helicity formalism%
}
\end{doublespace}

\vspace{4em}

{\large
	Jakub~Kuczmarski\footnote{%
		e-mail:
		\href{mailto:Jakub.Kuczmarski@fuw.edu.pl}{Jakub.Kuczmarski@fuw.edu.pl}%
	}
}

\vspace{1em}

\textit{%
	Institute of Theoretical Physics,
	Faculty of Physics,
	University of Warsaw%
}

\vspace{3em}

\textbf{Abstract}
\end{center}

\begin{onehalfspace}
	We present \texttt{SpinorsExtras} package implementing massive
	spinor-helicity formalism in Mathematica on top of \texttt{S@M}
	package.  Package defines new objects for Mathematica - massive
	spinors and reference, associated and polarization vectors. Various
	properties of introduced symbols and relations among them are
	implemented together with functions to manipulate and interchange such
	objects. The package can be used for symbolic calculation of helicity
	amplitudes involving massive particles with real or complex momenta, using
	on- or off-shell recursive techniques or Feynman diagrams.
\end{onehalfspace}

\end{titlepage}

\section{Introduction}
\label{sec:introduction}

The Standard Model (SM) of strong and electroweak interactions has
been successfully tested in numerous experiments. With very few
exceptions, which still require further confirmation, results of them
agree, within the experimental and theoretical errors, with the
predictions of SM (extended with massive neutrino sector).

The constantly increasing precision of various measurements require
relevant progress in accuracy of theoretical computations of SM
expectations.  It is especially important in the LHC era. The new
energy range accessible in LHC experiments gives hope to finally spot
some signs on New Physics - either direct new particle production or
deviation from SM predictions in observables related to production of
already known particles. However, the increase of achievable energy in
hadron collider like LHC leads also to very serious complications in
both data analysis and theoretical calculations. The latter include,
in particular, calculating very tedious multi-leg and/or multi-loop
QCD corrections to various amplitudes and proton PDF functions.  The
significant progress in simplifying and automatizing QCD calculations
has been achieved in recent years using the spinor-helicity formalism
and on-shell recursion techniques.

Spinor helicity formalism~\cite{Berends:1981rb, DeCausmaecker:1981bg,
  Berends:1981uq, Caffo:1982ds, Gunion:1985vca, Kleiss:1985yh,
  Xu:1986xb} consist of expressing scattering amplitudes calculated in
quantum field theory by chiral spinors, objects that are transforming
in lowest dimensional non-trivial representations of Lorentz
group. Those ``smallest pieces'' give very granular control over
amplitude and high potential for simplifications while retaining
explicit Lorentz invariance.

The formalism is a natural language for an ``on-shell recursion'' in
which only physical states propagate in amplitudes. It was first
derived for massless Yang-Mills theories \cite{Britto:2004ap,
  Britto:2005fq} and developed into practical methods and tools of
calculating QCD amplitudes.  However, LHC with its high energy range
and high luminosity is also an efficient source of massive particles -
heavy quarks, massive gauge bosons and Higgs particles. To analyze
their production, decays or more complicated processes (like for
example potential unitarity violation in massive gauge boson
rescattering) the generalization of on-shell recursion for theories
with massive particles is necessary. Such generalizations have been
discussed in several papers~\cite{Badger:2005zh, Badger:2005jv,
  Ozeren:2006ft, Schwinn:2007ee, Hall:2007mz, Cheung:2008dn,
  Cohen:2010mi, Chen:2011sba, Britto:2012qi}.

Efficient recursive calculation of complicated amplitudes, especially
with many external particles, requires large degree of automatization
and extensive use of computers for both analytical and numerical
computations.  To facilitate them, various packages and libraries of
routines have been developed and published for general
use~\cite{Gleisberg:2008fv, Berger:2008sj, Giele:2008bc,
  Lazopoulos:2008ex, Giele:2009ui, Cullen:2010jv, Dixon:2010ik,
  Bourjaily:2010wh, Badger:2010nx, Bevilacqua:2011xh, Hirschi:2011pa,
  Cullen:2011ac}.  In particular, set of routines designed for
manipulating objects used in the massless spinor-helicity formalism in
Mathematica has been implemented in the \texttt{S@M} (``Spinors at
Mathematica'') package \cite{Maitre:2007jq}.  \texttt{SpinorsExtras}
package presented in this article extends \texttt{S@M} capabilities by
introducing symbols representing states of massive fermions: massive
spinors, and vector bosons: polarization vectors along with their
properties, relations between them and functions to manipulate and
interchange them within the Mathematica code. Package provides also
tools for management of reference vectors that determine the
quantization axes, proportionality relation among spinors and various
phase conventions for spinors and polarization vectors.

Package is well suited for calculation of helicity amplitudes
involving massive particles using recursive (including on-shell)
methods, Feynman diagrams, or other techniques employing massive
spinor helicity formalism, but some of provided tools are also useful
for easier manipulations of purely massless amplitudes.

The outline of this article is as follows. In Sec.~\ref{sec:notation}
we define the notation used in the massive spinor-helicity formalism,
with some extensions introduced (comparing to commonly used
conventions).  Sec.~\ref{sec:mathematica package} provides technical
details of package installation, bug reports,
etc. Sec.~\ref{sec:symbols reference} contains reference manual for
all public symbols provided by the package. In Sec.~\ref{sec:tutorial}
we present tutorial with simple example of package usage.

\section{Notation}
\label{sec:notation}

\subsection{Chiral (massless) spinors}

Dirac spinors for massless particle with momentum $k$ and definite
helicity/chirality are denoted by:
\begin{subequations}\label{massless spinors def}\begin{align}
	\label{massless spinors def ket}
	|k\rangle & = u^+(k) = v^-(k) &
	|k] & = u^-(k) = v^+(k)
	\\
	\label{massless spinors def bra}
	[k| & = \bar u^+(k) = \bar v^-(k) &
	\langle k| & = \bar u^-(k) = \bar v^+(k)
\end{align}\end{subequations}

For chiral spinors and their products we use conventions from
\texttt{S@M} package \cite{Maitre:2007jq}. Following this conventions
angle spinors will also be called A spinors and square spinors will be
also called B spinors.

\subsection{Light cone decomposition}

\subsubsection{Lightlike reference vector}

With any non-lightlike four-vector $k$, we can associate lightlike
four-vector $k^q$ using light cone decomposition with respect to an
arbitrary lightlike four-vector $q$ (such that $k \cdot q \neq 0$)
\cite{Kleiss:1985yh, Kosower:2004yz}:
\begin{equation}\label{light cone decomp def}
	k^q = k - \frac{k^2}{2k \cdot q} q
\end{equation}
In the literature one can find various conventions for denoting
associated vectors. \texttt{SpinorsExtras} package supports two conventions:
$k^q$ defined above which we call associated vector with
explicit reference vector and $k^\flat$ which we call associated
vector with implicit reference vector.

Whenever \texttt{SpinorsExtras} package needs to infer explicit
reference vector from associated vector with implicit reference vector
$k^\flat$, ``default reference vector for $k$'', denoted by $q_k$, is
used. In numerical considerations for $k^\mu = (k^0, \vec k)$ default
reference vector has following components $q_k = (\pm |\vec k|, -\vec
k)$ with minus for real negative $k^0$ and plus when $k^0$ is real non-negative
or has nonzero imaginary part.

\subsubsection{Non-lightlike reference vector}

Given two non-proportional non-lightlike four-vectors $k$ and $p$
``simultaneous light cone decomposition'' can be performed such that
vector associated with one momentum is reference vector for the other
\cite{Schwinn:2007ee}
\begin{align}\label{light cone decomp simultaneous def eq}
	k & = k^p + \frac{k^2}{2k^p \cdot p^k} p^k &
	p & = p^k + \frac{p^2}{2p^k \cdot k^p} k^p
\end{align}
Solving above system of equations for $k^p$ and $p^k$ gives unique, up
to rescaling, lightlike basis of two-dimensional space, spanned by $k$
and $p$.
\begin{align}\label{light cone decomp simultaneous def explicit}
	k^p & = \frac{
		\left(\sgn(k \cdot p)\sqrt{\Delta} + k \cdot p\right) k - k^2 p
	}{2\sgn(k \cdot p)\sqrt{\Delta}}
	&
	p^k & = \frac{
		\left(\sgn(k \cdot p)\sqrt{\Delta} + k \cdot p\right) p - p^2 k
	}{2\sgn(k \cdot p)\sqrt{\Delta}}
\end{align}
where $\Delta = (k \cdot p)^2 - p^2 k^2$.  For $k^2 = 0$ ($p^2 = 0$)
formula for $p^k$ ($k^p$) reduces to \eqref{light cone decomp def}.
It's worth noting that for any vector $q$ (lightlike or not)
non-proportional to $k$, belonging to space spanned by $k$ and $p$ and
such that $k^p \cdot q \neq 0$, we have: $k^q = k^p$. In particular
$k^{p^k} = k^p$.

\subsection{Eigenstates of Pauli-Lubański projections}

In relativistic theories one can define Pauli-Lubański pseudovector
$W^\mu = \frac{1}{2} \epsilon^{\mu\nu\rho\sigma}J_{\nu\rho}k_\sigma$
where: $\epsilon_{\mu\nu\rho\sigma}$ is the Levi-Civita symbol,
$J_{\nu\rho}$ is the relativistic angular momentum tensor and
$k_\sigma$ is the four-momentum.

We will consider eigenstates of normalized projection of
Pauli-Lubański pseudovector $W$ to axis given by arbitrary lightlike
reference vector $q$ such that $k \cdot q \ne 0$ i.e. eigenstates of
$W(q)$ operator defined by:
\begin{equation}\label{Pauli-Lubański projection def}
	W(q) = \frac{W \cdot q}{k \cdot q}
\end{equation}

Multiplication of reference vector $q$ by any non-zero complex number
does not change $W(q)$ operator.

For any massive particle and any real lightlike reference vector $q$ there
exist reference frames in which three-momentum of this particle is parallel to
spatial part of reference vector: $\vec q$. $W(q)$ operator reduces to helicity
operator $\frac{\vec S \cdot \vec k}{|\vec k|}$ in those reference frames, so
eigenstates of $W(q)$ are also eigenstates of helicity in such frames. For
complex reference vector $q$ such reference frames don't have to exist.

Considering rest frame of arbitrary timelike four-vector $P$, one may ask for
which reference vector $q$ operator $W(q)$ reduces to helicity operator in this
reference frame. For particle with four-momentum $k$ ($P$ can not be
proportional to $k$) helicity operator in considered frame is given by $W(Q^k)$
for $Q$ being any linear combination of $P$ and $k$, that is not proportional
to $k$, e.g. $Q = P$ or $Q = P - k$.

\subsubsection{Massive Dirac spinors}

Dirac spinors associated with (possibly complex) four-vector $k$ that are
eigenstates of $W(q)$ are given by \cite{Kleiss:1985yh, Schwinn:2007ee,
Hall:2007mz}:
\begin{subequations}\label{massive spinors def}
	\begin{align}
		\label{massive spinors def u ket}
		|\prescript{q}{+}k\rangle &
			= u^+(k,q)
			= |k^q\rangle + \frac{m_k}{[k^q|q]}|q] &
		|\prescript{q}{+}k] &
			= u^-(k,q)
			= |k^q] + \frac{m_k}{\langle k^q|q\rangle}|q\rangle
		\\
		\label{massive spinors def u bra}
		[\prescript{q}{+}k| &
			= \bar u^+(k,q)
			= [k^q| + \frac{m_k}{\langle q|k^q\rangle}\langle q| &
		\langle \prescript{q}{+} k| &
			= \bar u^-(k,q)
			= \langle k^q| + \frac{m_k}{[q|k^q]}[q|
	\end{align}
	\begin{align}
		\label{massive spinors def v ket}
		|\prescript{q}{-}k\rangle &
			= v^-(k,q)
			= |k^q\rangle - \frac{m_k}{[k^q|q]}|q] &
		|\prescript{q}{-}k] &
			= v^+(k,q)
			= |k^q] - \frac{m_k}{\langle k^q|q\rangle}|q\rangle
		\\
		\label{massive spinors def v bra}
		[\prescript{q}{-}k| &
			= \bar v^-(k,q)
			= [k^q| - \frac{m_k}{\langle q|k^q\rangle}\langle q| &
		\langle \prescript{q}{-} k| &
			= \bar v^+(k,q)
			= \langle k^q| - \frac{m_k}{[q|k^q]}[q|
	\end{align}
\end{subequations}
Where by mass $m_k$ we denote the principal square root of (possibly complex)
$k^2$, i.e. $m_k = \sqrt{|k^2|}e^{i\frac{\varphi}{2}}$ for
$k^2 = |k^2|e^{i\varphi}$, $-\pi < \varphi \le \pi$.

\texttt{SpinorsExtras} package
\hyperlink{SpinorsExtras/guide/section/Massive}{%
	supports the convention above%
},
in which massive spinors are labeled by four-momentum.
Comparing to other conventions, we introduced left subscript $+$ or $-$ (which
we call mass sign, since it corresponds to two possible ``signs of mass'' such
that $(\pm m_k)^2 = k^2$) denoting $u$ or $v$ spinors respectively and left
superscript, if present, denoting explicit reference vector. If left
superscript is not present massive spinor is considered to have implicit
reference vector.

For $m_k = 0$ above spinors become independent of reference vector $q$
and reduce to \eqref{massless spinors def}.

It's worth noting that for any solution of massive Dirac equation
there always exist reference vector $q$ such that this solution takes
one of the forms presented above.

\subsubsection{Polarization vectors}

Transverse polarization vectors for particle with momentum $k$ and
mass $m_k$ that are eigenstates of $W(q)$ are given by
\cite{Kosower:2004yz}:
\begin{subequations}\label{pol vec def massive}
\begin{align}\label{pol vec def massive transverse}
	\epsilon^{+}_\mu(k,q) &
		= \frac{\langle q|\gamma_\mu|k^q]}{\sqrt{2} \langle q|k^q\rangle} &
	\epsilon^{-}_\mu(k,q) &
		= \frac{[q|\gamma_\mu|k^q\rangle}{\sqrt{2} [q|k^q]}
\end{align}

and longitudinal polarization vector is given by:
\begin{equation}\label{pol vec def massive longitudinal}
	\epsilon^{0}_\mu(k,q) = -\frac{k_\mu}{m_k} + \frac{m_k}{k \cdot q} q_\mu
		= -\frac{k^q_\mu}{m_k} + \frac{m_k}{2k^q \cdot q} q_\mu
\end{equation}

Considerations involving exchanges of off-shell vector bosons
sometimes require use of ``scalar polarization'' which is just
rescaled momentum:
\begin{equation}\label{pol vec def massive scalar}
	\epsilon^{\text{S}}_\mu(k,q) = \frac{k_\mu}{m_k}
\end{equation}
\end{subequations}

Polarization vectors of massless particles with momentum $k$,
transverse to subspace spanned by $k$ and arbitrary lightlike
reference vector $q$ can be taken the same as transverse vectors for
massive case \eqref{pol vec def massive}, but in this case $k$ is
lightlike by itself so $k^q = k$ \cite{Gunion:1985vca, Kleiss:1985yh,
  Xu:1986xb, Kosower:2004yz}. It's worth noting that in massless case
difference between polarization vectors for different reference
vectors $q$ is proportional to momentum $k$ \cite{Kleiss:1985yh,
  Xu:1986xb}.

\texttt{SpinorsExtras} package introduces
\hyperlink{SpinorsExtras/guide/section/Pol}{%
	labels for above polarization vectors and functions for manipulating them%
}.
Each polarization vector has also variant with implicit reference vector.

\subsection{Spinor chains}

\texttt{SpinorsExtras} package extends capabilities of \texttt{S@M} package by
adding support for spinor chains involving massive spinors.

Spinor chains in our convention are denoted by:
\begin{align}\label{spinor chains def}
	\langle \prescript{q_1}{+} k_1 | k_2 | \ldots | k_{n-1} |
		\prescript{q_n}{+}k_n \rangle
	& = \bar u^-(k_1,q_1) \slashed{k}_2 \ldots \slashed{k}_{n-1} u^+(k_n,q_n)
\end{align}
and analogously, using \eqref{massive spinors def}, for square bracket at one
or both sides and for chains involving $v$ spinors.

Spinor chains with angle brackets on both ends are antisymmetric, same is
true for square brackets:
\begin{subequations}\label{spinor chains antisymmetry}
\begin{align}
	\label{spinor chains antisymmetry Spaa}
	\langle \prescript{q_1}{\pm} k_1 | k_2 | \ldots| k_{n-1} |
		\prescript{q_n}{\pm}k_n \rangle
	& = - \langle \prescript{q_n}{\pm} k_n | k_{n-1} | \ldots| k_2 |
		\prescript{q_1}{\pm}k_1 \rangle
	\\
	\label{spinor chains antisymmetry Spbb}
	[ \prescript{q_1}{\pm} k_1 | k_2 | \ldots| k_{n-1} |
		\prescript{q_n}{\pm}k_n ]
	& = - [ \prescript{q_n}{\pm} k_n | k_{n-1} | \ldots| k_2 |
		\prescript{q_1}{\pm}k_1 ]
\end{align}
\end{subequations}

Spinor chains with different types of brackets on ends are symmetric:
\begin{align}\label{spinor chains symmetry}
	\langle \prescript{q_1}{\pm} k_1 | k_2 | \ldots| k_{n-1} |
		\prescript{q_n}{\pm}k_n ]
	& = [ \prescript{q_n}{\pm} k_n | k_{n-1} | \ldots| k_2 |
		\prescript{q_1}{\pm}k_1 \rangle
\end{align}

Above identities are implemented in \texttt{SpinorsExtras} package.

\subsection{Composite four-vectors}

Any, lightlike or non-lightlike, four-vector $k$ can be expressed by spinors in
following way:
\begin{equation}\label{composite vectors}
	k^\mu = \frac{1}{2}
		[ \prescript{q}{\pm} k | \gamma^\mu | \prescript{q}{\pm}k \rangle
\end{equation}
Reference vectors and mass signs of both spinors are the same. For
real momentum and real reference vector angle and square spinors are
Dirac conjugations of each others.

In many calculations it is useful to represent complex four-vector as a
composition of two spinors, such that each of them is related to different real
four-vector, that is also present in given calculation (like
e.g. transverse polarization vectors or vector used for performing
BCFW shifts). For such situations \texttt{SpinorsExtras} package
\hyperlink{SpinorsExtras/guide/section/Composite}{%
	introduces general composite vector labels%
}
$\eta(b,a)$, such that:
\begin{equation}\label{composite vector}
	\eta^\mu(b,a) = \frac{1}{2} [ b | \gamma^\mu | a \rangle
\end{equation}
where $b$ and $a$ are arbitrary massless or massive spinor labels.

When both spinors are massless then $\eta(b,a)$ is also massless and we have:
\begin{align}\label{composite vector both massless}
	\slashed{\eta}(b,a) & = | b ] \langle a | + | a \rangle [ b | &
	[ \eta(b,a) | & = [ b | &
	| \eta(b,a) \rangle & = | a \rangle
\end{align}

When one of spinors is massless and second is massive, composite
vector is still massless:
\begin{subequations}\label{composite vector one massive}
\begin{align}
	\label{composite vector one massive: left}
	\eta^\mu(\prescript{}{\pm} k,a) &
		= \frac{1}{2} [ \prescript{}{\pm} k | \gamma^\mu | a \rangle
		= \frac{1}{2} [ k^\flat | \gamma^\mu | a \rangle
		= \eta^\mu(k^\flat,a) &
	[ \eta(\prescript{}{\pm} k,a) | & = [ k^\flat | &
	| \eta(\prescript{}{\pm} k,a) \rangle & = | a \rangle
	\\
	\label{composite vector one massive: right}
	\eta^\mu(b,\prescript{}{\pm} k) &
		= \frac{1}{2} [ b | \gamma^\mu | \prescript{}{\pm} k \rangle
		= \frac{1}{2} [ b | \gamma^\mu | k^\flat \rangle
		= \eta^\mu(b,k^q) &
	[ \eta(b,\prescript{}{\pm} k) | & = [ b | &
	| \eta(b,\prescript{}{\pm} k) \rangle & = | k^\flat \rangle
\end{align}
\end{subequations}

For both spinors massive, composite vector is massive:
\begin{equation}\label{composite vector both massive mass}
	\eta(\prescript{}{s_k} k,\prescript{}{s_p} p) \cdot
	\eta(\prescript{}{s_k} k,\prescript{}{s_p} p)
		= m_k m_p \frac{
				\langle q_k | p^\flat \rangle [ k^\flat | q_p ]
			}{
				\langle q_k | k^\flat \rangle [ p^\flat | q_p ]
			}
\end{equation}
and decomposes to sum of two massless vectors:
\begin{align}\label{composite vector both massive}
	\eta^\mu(\prescript{}{s_k} k,\prescript{}{s_p} p)
		= \frac{1}{2} [ \prescript{}{s_k} k | \gamma^\mu |
			\prescript{}{s_p} p \rangle
		&
		= \frac{1}{2} [ k^\flat | \gamma^\mu | p^\flat \rangle
			+ s_k s_p \frac{1}{2}
				\frac{m_k m_p}{\langle q_k | k^\flat \rangle [ p^\flat | q_p ]}
				\langle q_k | \gamma^\mu | q_p ]
	\\ &
		= \eta^\mu(k^\flat, p^\flat)
			+ s_k s_p
				\frac{m_k m_p}{\langle q_k | k^\flat \rangle [ p^\flat | q_p ]}
				\eta^\mu(q_k, q_p)
\end{align}
where $s_k, s_p = \pm 1$ are mass signs of spinors associated with $k$
and $p$ respectively.

\subsection{Proportional spinors and vectors}

Products of two massless angle spinors $| i \rangle$, $| j \rangle$ vanish
$\langle i | j\rangle = 0$ if and only if those spinors are proportional. The
same is true for square spinors. Thus proportionality is an important relation
among spinors which leads to simplifications and restrictions on possible
choices of reference vectors.

Two light-like four-vectors are proportional if and only if both spinors
related to one vector are proportional to corresponding spinors of other
vector.

\texttt{SpinorsExtras} package introduces
\hyperlink{SpinorsExtras/guide/section/Proportional}{
tools for establishing proportionality relations among spinors
}.

\subsection{Reference vectors}

Quantities like associated vectors, massive spinors and polarization
vectors depend on respective reference vectors. Amplitudes involving
massless vector bosons or squares of absolute values of amplitudes
with massive particles summed over spin states should be independent
of choice of reference vectors.

\texttt{SpinorsExtras} package introduces
\hyperlink{SpinorsExtras/guide/section/Ref}{
tools for checking this invariance and for choosing reference vectors
for which expression take simplest forms
}.

\subsection{BCFW shifts}

In addition to already implemented in \texttt{S@M} shifts of massless spinors,
\texttt{SpinorsExtras} implements shifts suitable for massive fermions
described in \cite{Schwinn:2007ee}. For two non-lightlike shifted momenta $p$
and $k$ the holomorphic shift is defined as:
\begin{subequations}\label{massive shift holomorphic}
\begin{align}
	\label{massive shift holomorphic p}
	p & \to p + z \eta(p^k,k^p) &
	| \prescript{k^p}{\pm}p \rangle & \to
		| \prescript{k^p}{\pm}p \rangle + z | k^p \rangle &
	[ \prescript{k^p}{\pm}p | & \to
		[ \prescript{k^p}{\pm}p |
	\\
	\label{massive shift holomorphic k}
	k & \to k - z \eta(p^k,k^p) &
	| \prescript{p^k}{\pm}k \rangle & \to
		| \prescript{p^k}{\pm}k \rangle &
	[ \prescript{p^k}{\pm}k | & \to
		[ \prescript{p^k}{\pm}k | - z [ p^k |
\end{align}
\end{subequations}
and anti-holomorphic shift as:
\begin{subequations}\label{massive shift anti-holomorphic}
\begin{align}
	\label{massive shift anti-holomorphic p}
	p & \to p + z \eta(k^p,p^k) &
	| \prescript{k^p}{\pm}p \rangle & \to
		| \prescript{k^p}{\pm}p \rangle &
	[ \prescript{k^p}{\pm}p | & \to
		[ \prescript{k^p}{\pm}p | + z [ k^p |
	\\
	\label{massive shift anti-holomorphic k}
	k & \to k - z \eta(k^p,p^k) &
	| \prescript{p^k}{\pm}k \rangle & \to
		| \prescript{p^k}{\pm}k \rangle - z | p^k \rangle &
	[ \prescript{p^k}{\pm}k | & \to
		[ \prescript{p^k}{\pm}k |
\end{align}
\end{subequations}

\section{Mathematica package}
\label{sec:mathematica package}

Package is available on its webpage:
\url{http://www.fuw.edu.pl/~jkuczm/SpinorsExtras/}\\
Source code is also hosted on GitHub:
\url{https://github.com/jkuczm/SpinorsExtras}.

\subsection{Installation}

To use \texttt{SpinorsExtras} you also need Mathematica with
\href{http://www.slac.stanford.edu/~maitreda/Spinors/}{\texttt{S@M} package}
\cite{Maitre:2007jq} installed.

To install \texttt{SpinorsExtras} package evaluate in Mathematica following
expression:
\begin{flushleft}
    \includegraphics{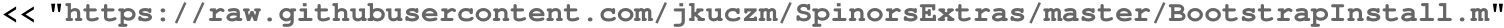}
\end{flushleft}

This will install
\href{https://github.com/lshifr/ProjectInstaller}{\texttt{ProjectInstaller}}
package, if you don't have it already installed. Then \texttt{SpinorsExtras}
will be installed together with following dependencies:
\href{https://github.com/jkuczm/MathematicaEvaluationUtilities}%
{\texttt{EvaluationUtilities}},
\href{https://github.com/jkuczm/MathematicaMessagesUtilities}%
{\texttt{MessagesUtilities}},\\
\href{https://github.com/jkuczm/MathematicaOptionsUtilities}%
{\texttt{OptionsUtilities}},
\href{https://github.com/jkuczm/MathematicaPatternUtilities}%
{\texttt{PatternUtilities}},
\href{https://github.com/jkuczm/MathematicaProtectionUtilities}%
{\texttt{ProtectionUtilities}},
\href{https://github.com/jkuczm/MathematicaStringUtilities}%
{\texttt{StringUtilities}}
and
\href{https://github.com/jkuczm/MUnitExtras}{\texttt{MUnitExtras}}.

You can install \texttt{SpinorsExtras} package manually by downloading\\
\href{http://www.fuw.edu.pl/~jkuczm/SpinorsExtras/SpinorsExtrasWithDependencies.zip}%
{\texttt{SpinorsExtrasWithDependencies.zip}}
and extracting it to any directory which is on Mathematica path. To see list of those directories evaluate
\lstinline|$Path| in Mathematica. Most appropriate directory is the one
starting with value of \lstinline|$UserBaseDirectory| variable with
\texttt{Applications} directory at the end.

To load the package evaluate \lstinline|Needs["SpinorsExtras`"]|

\subsection{Documentation}

\texttt{SpinorsExtras} package comes with documentation integrated
with Mathematica documentation center. To use it, open the
documentation center (\texttt{Help > Documentation Center}) and type
``SpinorsExtras'' in the search field.

Documentation for any individual symbol from \texttt{SpinorsExtras}
package, can be also called by selecting this symbols name in
Mathematica and pressing the F1 key.  Symbols reference pages in
documentation center, in addition to reference provided in this paper,
contain also many examples how to use given symbol, command or
function.

\href{http://www.fuw.edu.pl/~jkuczm/SpinorsExtras/reference/guide/SpinorsExtras.html}%
{Online version of documentation} is available on project website.

\subsection{Tests}

\texttt{SpinorsExtras} package is distributed with an extensive
automated test suite consisting of more than four thousand
tests. To run the tests one needs
\href{%
	http://reference.wolfram.com/workbench/index.jsp?topic=%
	/com.wolfram.eclipse.help/html/tasks/tester/tester.html%
}{\texttt{MUnit}}
package that is distributed with Wolfram Workbench.

\subsection{Bugs and requests}

Bug reports and feature requests can be posted as
\href{https://github.com/jkuczm/SpinorsExtras/issues}{GitHub issues}
(the preferred way) or sent directly to
\href{mailto:Jakub.Kuczmarski@fuw.edu.pl}{author's email}.

\section{Symbols reference}
\label{sec:symbols reference}

An overview of all public symbols introduced by SpinorsExtras package.

\begin{itemize}
    \item[]
    \hypertarget{SpinorsExtras/guide/section/Ref}{\textbf{Reference vectors and spinors}}
    \begin{itemize}
        \item[]
        \hyperref[SpinorsExtras/ref/SpRef]{SpRef}
        - Labels default reference spinor for given Lorentz vector.
        \item[]
        \hyperref[SpinorsExtras/ref/RefInvariantQ]{RefInvariantQ}
        - Tests whether expression is invariant with respect to changes of reference spinors.
        \item[]
        \hyperref[SpinorsExtras/ref/RefSimplify]{RefSimplify}
        - Finds simplest form of expression by inserting possible reference spinors.
        \item[]
        \hyperref[SpinorsExtras/ref/ExplicitRef]{ExplicitRef}
        - Changes implicit reference vectors, in given expressions, to explicit  reference vectors.
        \item[]
        \hyperref[SpinorsExtras/ref/ImplicitRef]{ImplicitRef}
        - Changes explicit reference vectors, in given expressions, to implicit  reference vectors.
    \end{itemize}
    \item[]
    \hypertarget{SpinorsExtras/guide/section/Massive}{\textbf{Massive vectors and spinors}}
    \begin{itemize}
        \item[]
        \hyperref[SpinorsExtras/ref/SpM]{SpM}
        - Labels spinor for given massive momentum.
        \item[]
        \hyperref[SpinorsExtras/ref/SpAssoc]{SpAssoc}
        - Labels vector associated, by light cone decomposition, with massive momentum.
        \item[]
        \hyperref[SpinorsExtras/ref/LightConeDecompose]{LightConeDecompose}
        - Performs light cone decomposition of vectors and spinors.
        \item[]
        \hyperref[SpinorsExtras/ref/MassiveSpinorQ]{MassiveSpinorQ}
        - Tests whether given expression is interpretable as massive spinor.
        \item[]
        \hyperref[SpinorsExtras/ref/MassiveLVectorQ]{MassiveLVectorQ}
        - Tests whether given expression is interpretable as massive LVector.
    \end{itemize}
    \item[]
    \hypertarget{SpinorsExtras/guide/section/Composite}{\textbf{Composite vectors}}
    \begin{itemize}
        \item[]
        \hyperref[SpinorsExtras/ref/LvBA]{LvBA}
        - Labels vector composed of two independent spinors with different labels.
    \end{itemize}
    \item[]
    \hypertarget{SpinorsExtras/guide/section/Pol}{\textbf{Polarization vectors}}
    \begin{itemize}
        \item[]
        \hyperref[SpinorsExtras/ref/PolVec]{PolVec}
        - Labels polarization vector for given momentum, polarization and reference vector.
        \item[]
        \hyperref[SpinorsExtras/ref/ExpandPolVec]{ExpandPolVec}
        - Expresses polarization vectors by momentum and reference vectors.
        \item[]
        \hyperref[SpinorsExtras/ref/DeclarePossiblePol]{DeclarePossiblePol}
        - Sets given symbols to be treated as vector boson polarization.
        \item[]
        \hyperref[SpinorsExtras/ref/UndeclarePossiblePol]{UndeclarePossiblePol}
        - Removes given symbols from list of vector boson polarizations.
        \item[]
        \hyperref[SpinorsExtras/ref/PossiblePolQ]{PossiblePolQ}
        - Tests whether given expression is interpretable as vector boson polarization.
    \end{itemize}
    \item[]
    \hypertarget{SpinorsExtras/guide/section/Utilities}{\textbf{Utilities}}
    \begin{itemize}
        \item[]
        \hyperref[SpinorsExtras/ref/ReplaceLVector]{ReplaceLVector}
        - Replaces given Lorentz vector in given expression.
        \item[]
        \hyperref[SpinorsExtras/ref/ReplaceBSpinor]{ReplaceBSpinor}
        - Replaces given massless or massive B spinor in given expression.
        \item[]
        \hyperref[SpinorsExtras/ref/ReplaceASpinor]{ReplaceASpinor}
        - Replaces given massless or massive A spinor in given expression.
        \item[]
        \hyperref[SpinorsExtras/ref/ExpandMPToSpinors]{ExpandMPToSpinors}
        - Replaces Minkowski products by spinor products.
        \item[]
        \hyperref[SpinorsExtras/ref/ExpandSToMPs]{ExpandSToMPs}
        - Replaces s invariants by Minkowski products.
        \item[]
        \hyperref[SpinorsExtras/ref/DeclarePlusMinusOne]{DeclarePlusMinusOne}
        - Sets given symbols to be treated as $\pm $1.
        \item[]
        \hyperref[SpinorsExtras/ref/UndeclarePlusMinusOne]{UndeclarePlusMinusOne}
        - Removes given symbols from list of expressions treated as $\pm $1.
        \item[]
        \hyperref[SpinorsExtras/ref/PlusMinusOneQ]{PlusMinusOneQ}
        - Tests whether given expression is interpretable as $\pm $1.
        \item[]
        \hyperref[SpinorsExtras/ref/AnySpinorQ]{AnySpinorQ}
        - Tests whether given expression is interpretable as massless or massive spinor.
    \end{itemize}
    \item[]
    \hypertarget{SpinorsExtras/guide/section/Proportional}{\textbf{Proportional spinors and vectors}}
    \begin{itemize}
        \item[]
        \hyperref[SpinorsExtras/ref/DeclareBSpinorProportional]{DeclareBSpinorProportional}
        - Declares that B spinors with given labels are proportional.
        \item[]
        \hyperref[SpinorsExtras/ref/DeclareASpinorProportional]{DeclareASpinorProportional}
        - Declares that A spinors with given labels are proportional.
        \item[]
        \hyperref[SpinorsExtras/ref/DeclareLVectorProportional]{DeclareLVectorProportional}
        - Declares that LVectors with given labels are proportional.
        \item[]
        \hyperref[SpinorsExtras/ref/BSpinorProportionalQ]{BSpinorProportionalQ}
        - Tests whether B spinors with given labels are proportional.
        \item[]
        \hyperref[SpinorsExtras/ref/ASpinorProportionalQ]{ASpinorProportionalQ}
        - Tests whether A spinors with given labels are proportional.
        \item[]
        \hyperref[SpinorsExtras/ref/LVectorProportionalQ]{LVectorProportionalQ}
        - Tests whether LVectors with given labels are proportional.
    \end{itemize}
    \item[]
    \hypertarget{SpinorsExtras/guide/section/Phase}{\textbf{Phases management}}
    \begin{itemize}
        \item[]
        \hyperref[SpinorsExtras/ref/AppendPhase]{AppendPhase}
        - Multiplies parts of expression with additional phases.
        \item[]
        \hyperref[SpinorsExtras/ref/Phase]{Phase}
        - Represents additional phase of given expression.
    \end{itemize}
    \item[]
    \hypertarget{SpinorsExtras/guide/section/Decompose}{\textbf{Spinor decomposition}}
    \begin{itemize}
        \item[]
        \hyperref[SpinorsExtras/ref/DecomposeBSpinor]{DecomposeBSpinor}
        - Decomposes B spinor in given basis.
        \item[]
        \hyperref[SpinorsExtras/ref/DecomposeASpinor]{DecomposeASpinor}
        - Decomposes A spinor in given basis.
    \end{itemize}
    \item[]
    \hypertarget{SpinorsExtras/guide/section/SimpleTensor}{\textbf{Simple Tensors}}
    \begin{itemize}
        \item[]
        \hyperref[SpinorsExtras/ref/SimpleTensorQ]{SimpleTensorQ}
        - Tests whether given expression represents simple tensor.
        \item[]
        \hyperref[SpinorsExtras/ref/SimpleTensorGetBLabel]{SimpleTensorGetBLabel}
        - Extracts B spinor from tensor product of B and A spinors.
        \item[]
        \hyperref[SpinorsExtras/ref/SimpleTensorGetALabel]{SimpleTensorGetALabel}
        - Extracts A spinor from tensor product of B and A spinors.
    \end{itemize}
    \item[]
    \hypertarget{SpinorsExtras/guide/section/Numerics}{\textbf{Numerics}}
    \begin{itemize}
        \item[]
        \hyperref[SpinorsExtras/ref/DeclareSpinorRandomMomentum]{DeclareSpinorRandomMomentum}
        - Generates random numerics for given spinor.
        \item[]
        \hyperref[SpinorsExtras/ref/GenComplexMomenta]{GenComplexMomenta}
        - Generates random complex momenta for spinors so that they sum to zero.
    \end{itemize}
    \item[]
    \hypertarget{SpinorsExtras/guide/section/SatMModifications}{\textbf{Functions from original Spinors` context with modified behavior}}
    \begin{itemize}
        \item[]
        \hyperref[SpinorsExtras/ref/SpOpen]{SpOpen}
        - Decomposes spinor chains to products of smaller spinor chains.
        \item[]
        \hyperref[SpinorsExtras/ref/ExpandSToSpinors]{ExpandSToSpinors}
        - Converts s invariants to products of spinor chains.
    \end{itemize}
\end{itemize}

\subsection{Reference vectors and spinors}

\subsubsection{SpRef}

\label{SpinorsExtras/ref/SpRef}

Labels default reference spinor for given Lorentz vector.

\begin{flushleft}
    \includegraphics{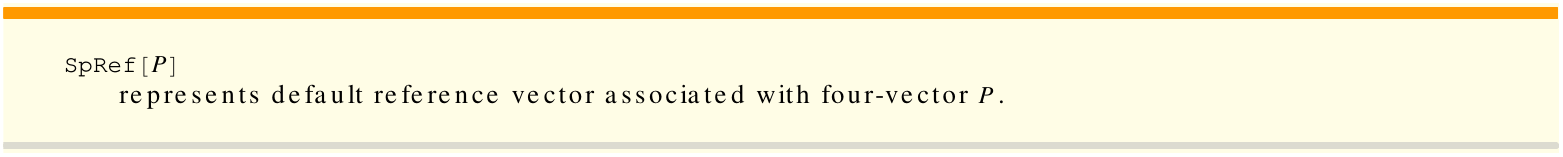}
    
    \includegraphics{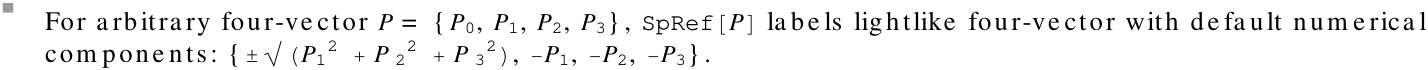}
\end{flushleft}

\subsubsection{RefInvariantQ}

\label{SpinorsExtras/ref/RefInvariantQ}

Tests whether expression is invariant with respect to changes of reference spinors.

\begin{flushleft}
    \includegraphics{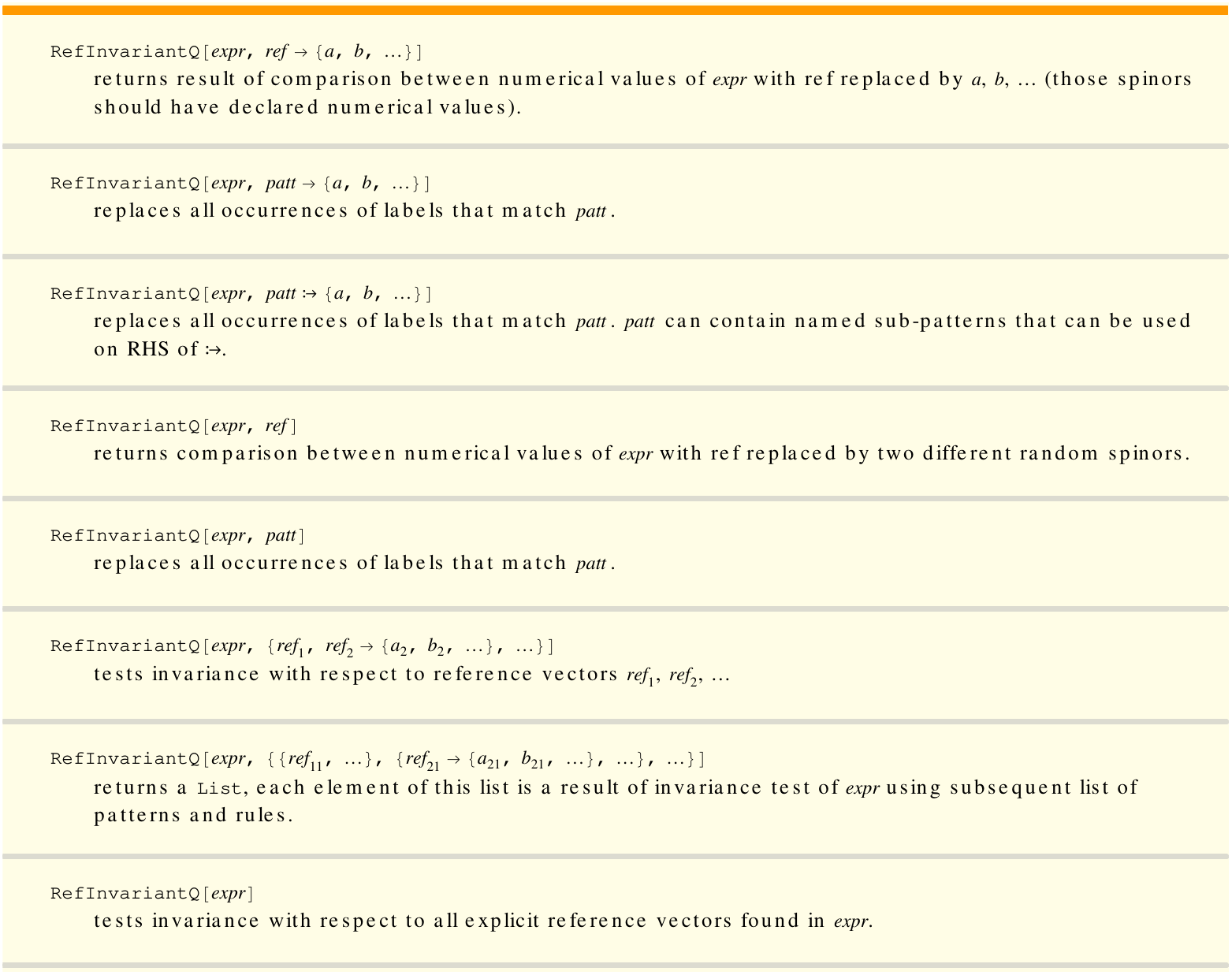}
    
    \includegraphics{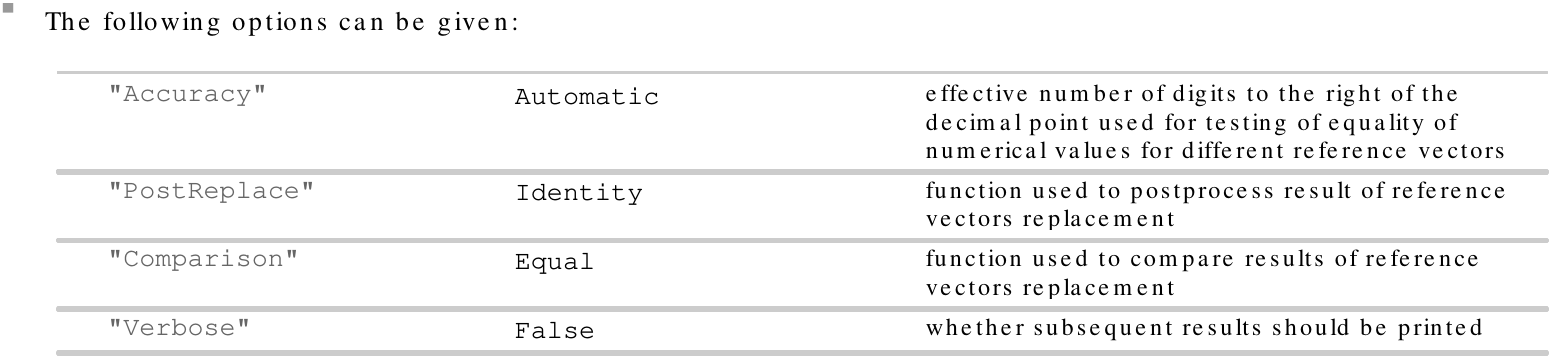}
\end{flushleft}

\subsubsection{RefSimplify}

\label{SpinorsExtras/ref/RefSimplify}

Finds simplest form of expression by inserting possible reference spinors.

\begin{flushleft}
    \includegraphics{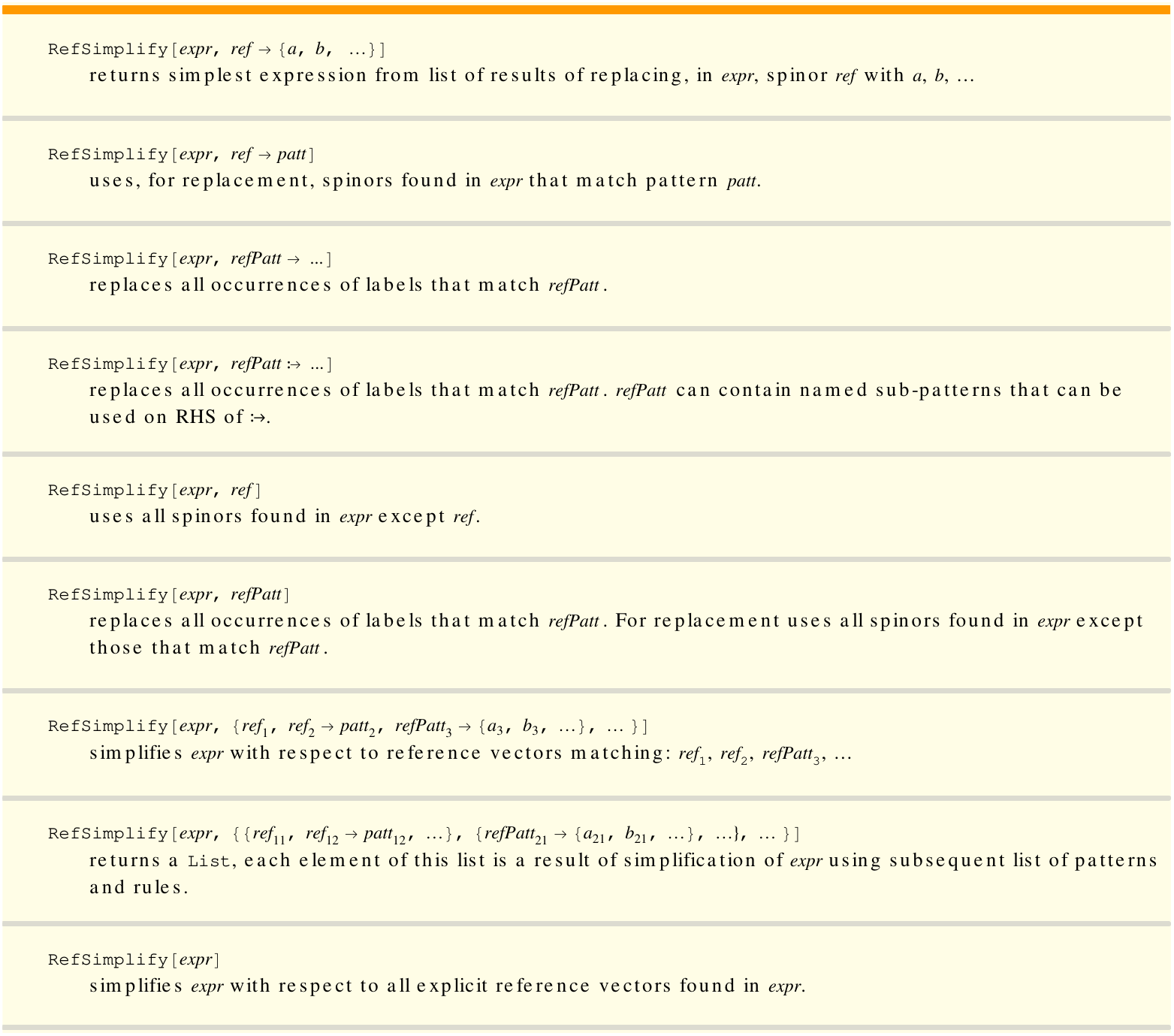}
    
    \includegraphics{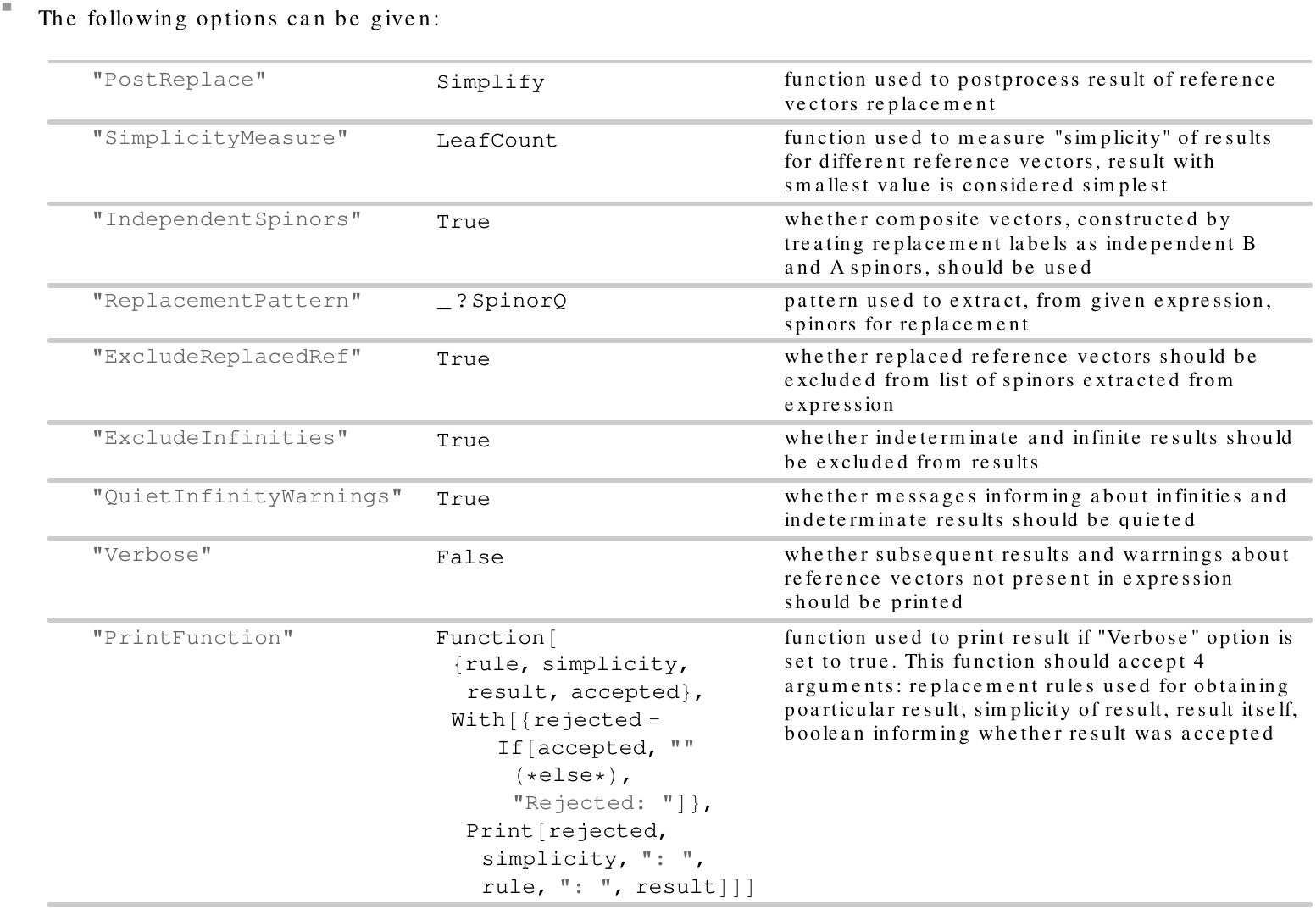}
\end{flushleft}

\subsubsection{ExplicitRef}

\label{SpinorsExtras/ref/ExplicitRef}

Changes implicit reference vectors, in given expressions, to explicit  reference vectors.

\begin{flushleft}
    \includegraphics{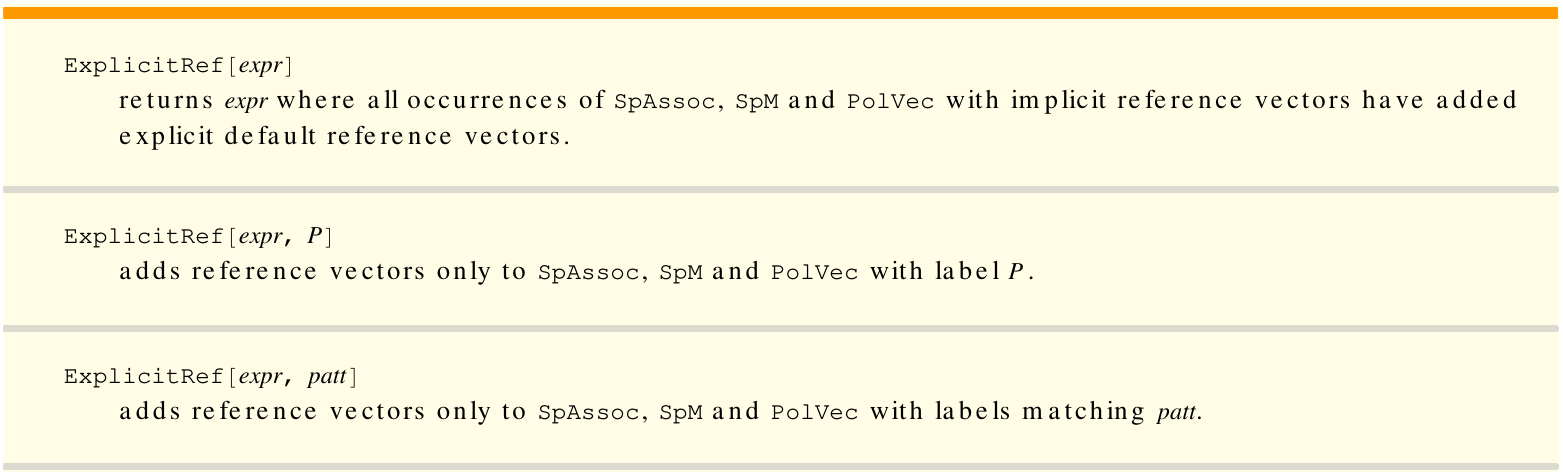}
\end{flushleft}

\subsubsection{ImplicitRef}

\label{SpinorsExtras/ref/ImplicitRef}

Changes explicit reference vectors, in given expressions, to implicit  reference vectors.

\begin{flushleft}
    \includegraphics{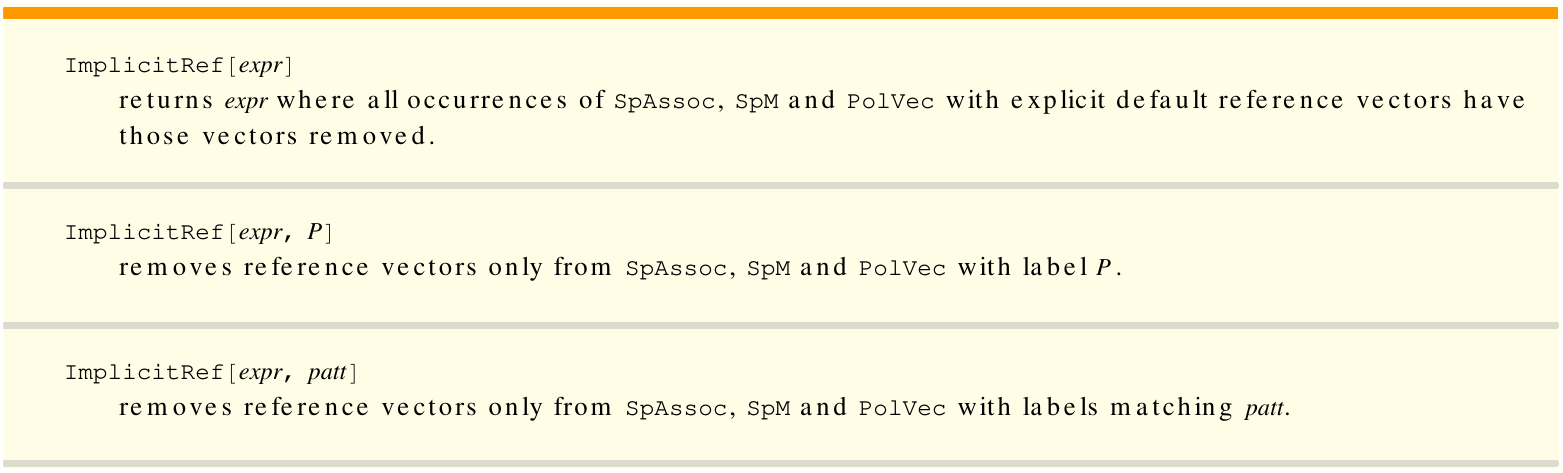}
    
    \includegraphics{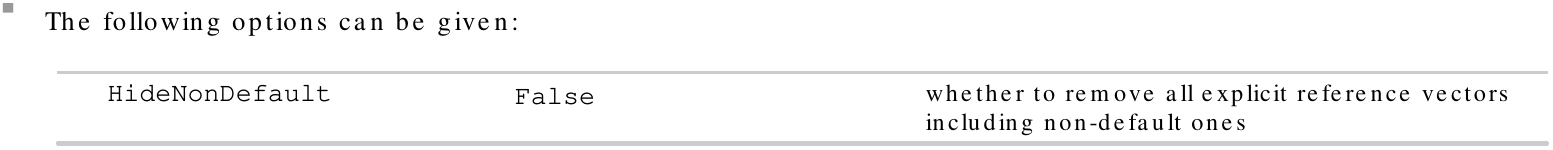}
\end{flushleft}

\subsection{Massive vectors and spinors}

\subsubsection{SpM}

\label{SpinorsExtras/ref/SpM}

Labels spinor for given massive momentum.

\begin{flushleft}
    \includegraphics{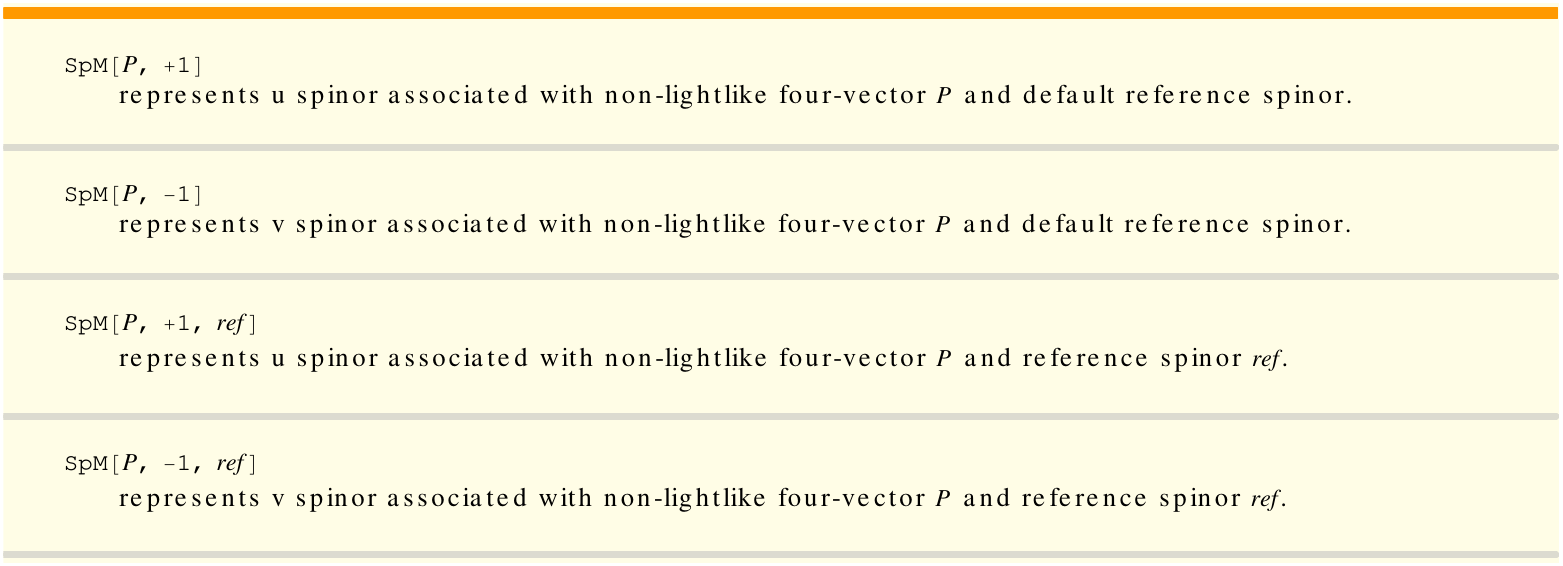}
\end{flushleft}

\subsubsection{SpAssoc}

\label{SpinorsExtras/ref/SpAssoc}

Labels vector associated, by light cone decomposition, with massive momentum.

\begin{flushleft}
    \includegraphics{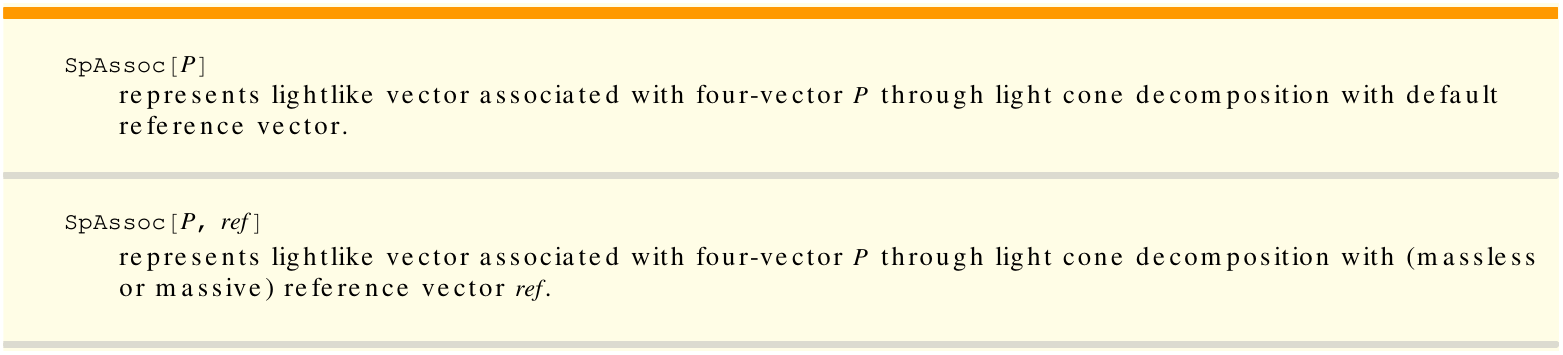}
    
    \includegraphics{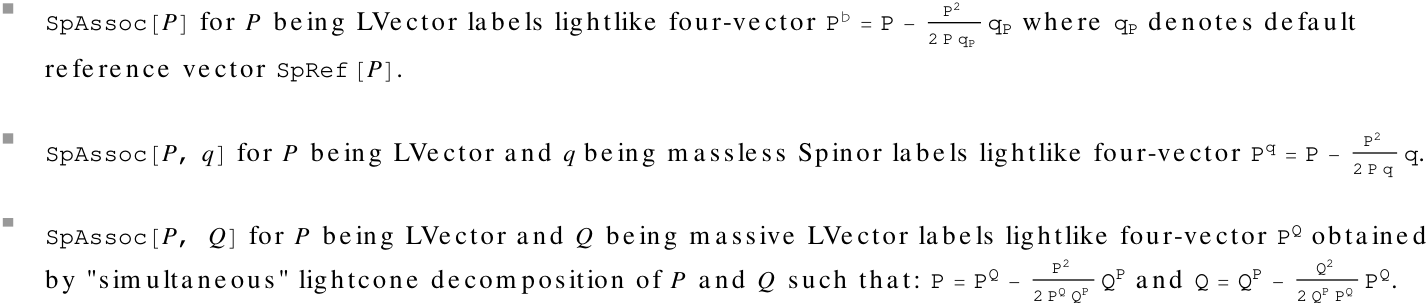}
\end{flushleft}

\subsubsection{LightConeDecompose}

\label{SpinorsExtras/ref/LightConeDecompose}

Performs light cone decomposition of vectors and spinors.

\begin{flushleft}
    \includegraphics{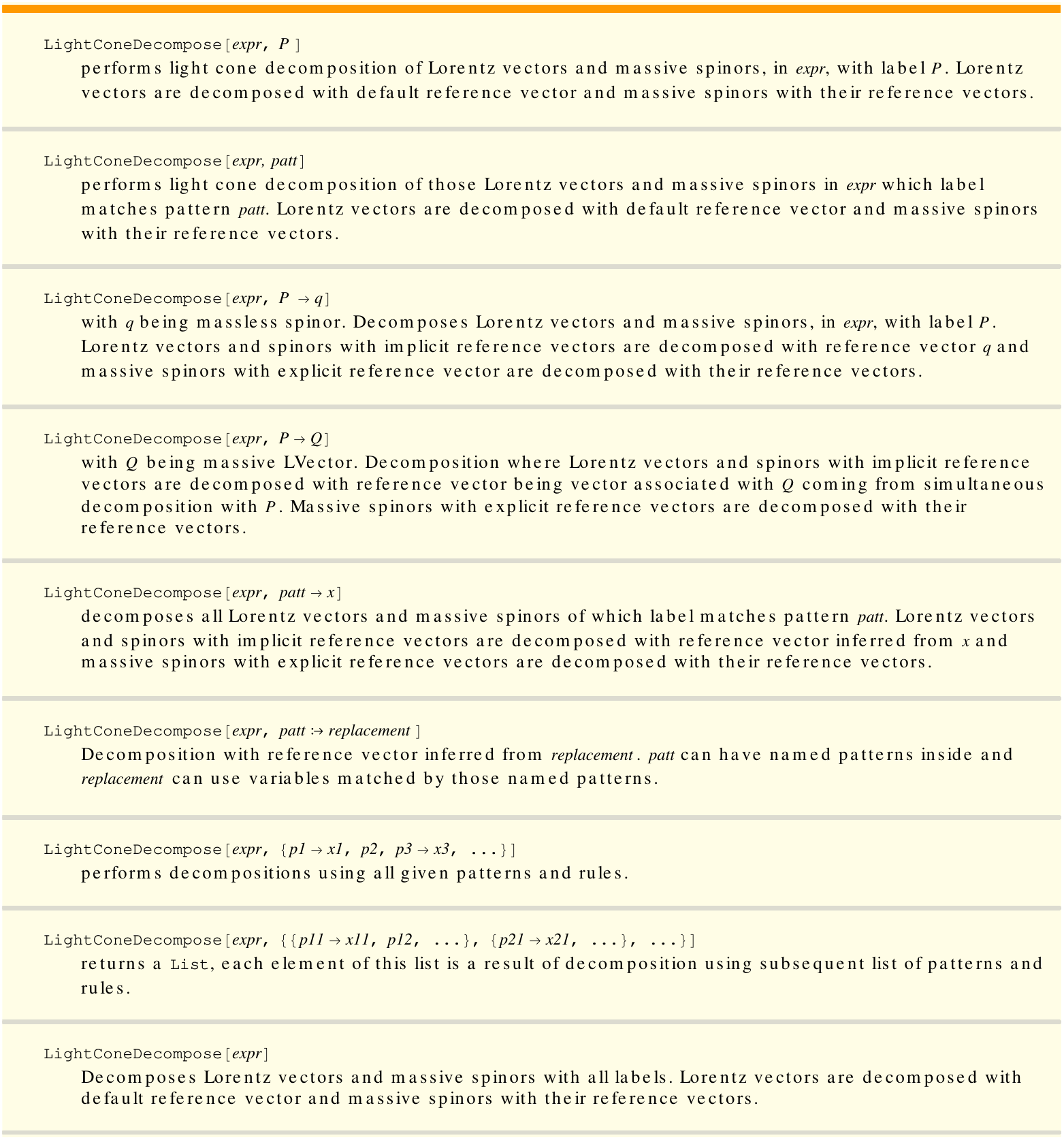}
    
    \includegraphics{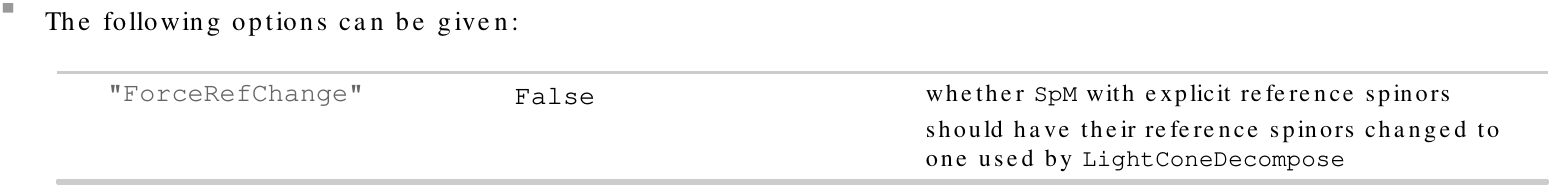}
\end{flushleft}

\subsubsection{MassiveSpinorQ}

\label{SpinorsExtras/ref/MassiveSpinorQ}

Tests whether given expression is interpretable as massive spinor.

\begin{flushleft}
    \includegraphics{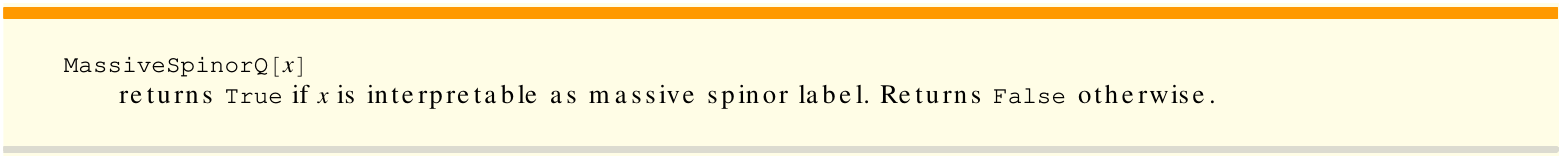}
\end{flushleft}

\subsubsection{MassiveLVectorQ}

\label{SpinorsExtras/ref/MassiveLVectorQ}

Tests whether given expression is interpretable as massive LVector.

\begin{flushleft}
    \includegraphics{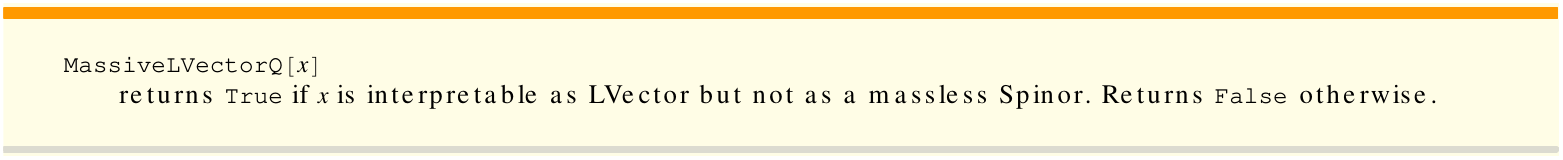}
\end{flushleft}

\subsection{Composite vectors}

\subsubsection{LvBA}

\label{SpinorsExtras/ref/LvBA}

Labels vector composed of two independent spinors with different labels.

\begin{flushleft}
    \includegraphics{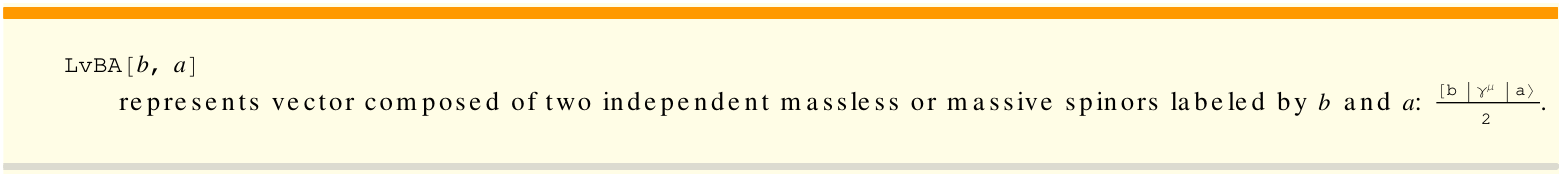}
    
    \includegraphics{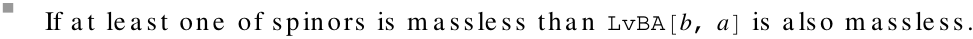}
\end{flushleft}

\subsection{Polarization vectors}

\subsubsection{PolVec}

\label{SpinorsExtras/ref/PolVec}

Labels polarization vector for given momentum, polarization and reference vector.

\begin{flushleft}
    \includegraphics{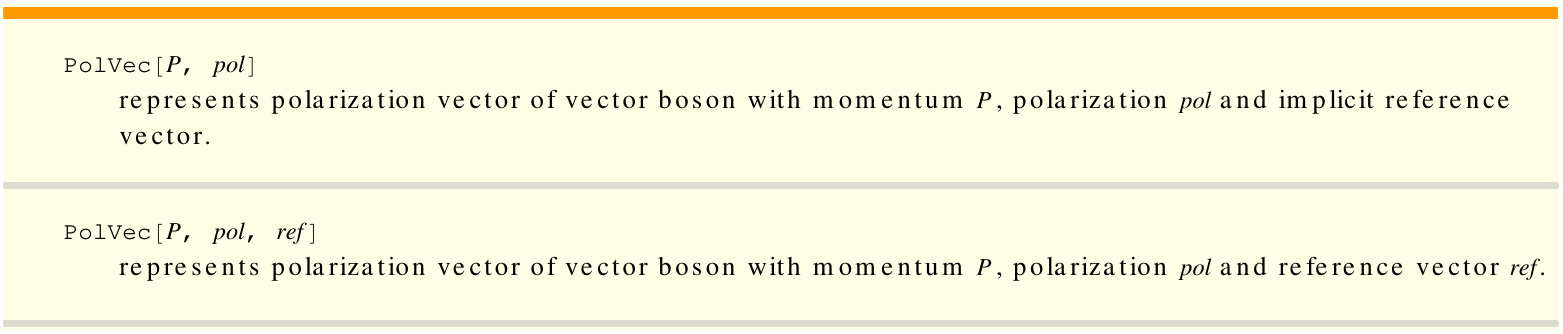}
    
    \includegraphics{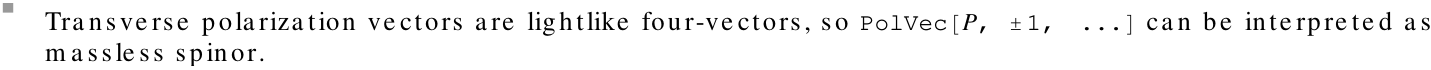}
\end{flushleft}

\subsubsection{ExpandPolVec}

\label{SpinorsExtras/ref/ExpandPolVec}

Expresses polarization vectors by momentum and reference vectors.

\begin{flushleft}
    \includegraphics{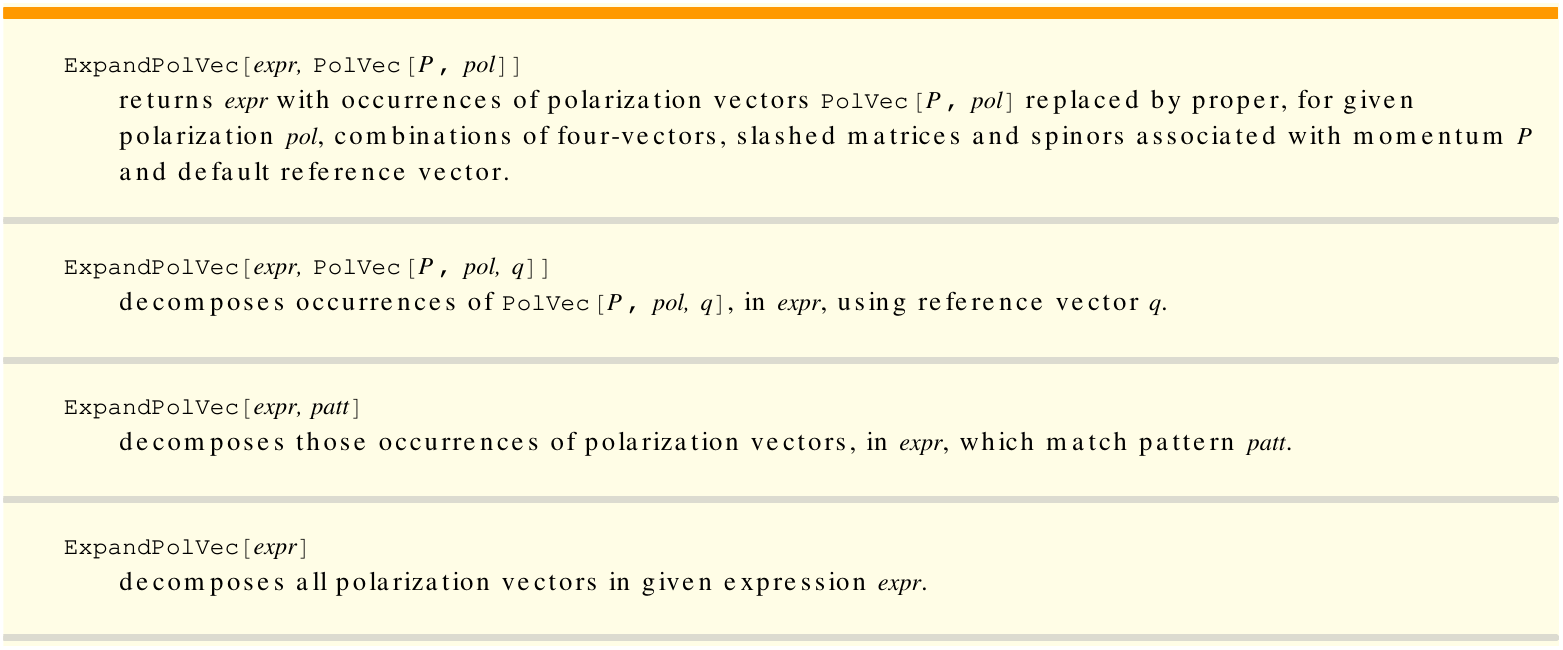}
    
    \includegraphics{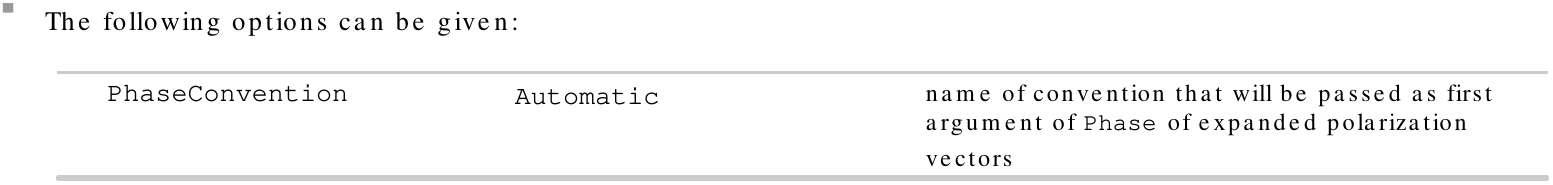}
\end{flushleft}

\subsubsection{DeclarePossiblePol}

\label{SpinorsExtras/ref/DeclarePossiblePol}

Sets given symbols to be treated as vector boson polarization.

\begin{flushleft}
    \includegraphics{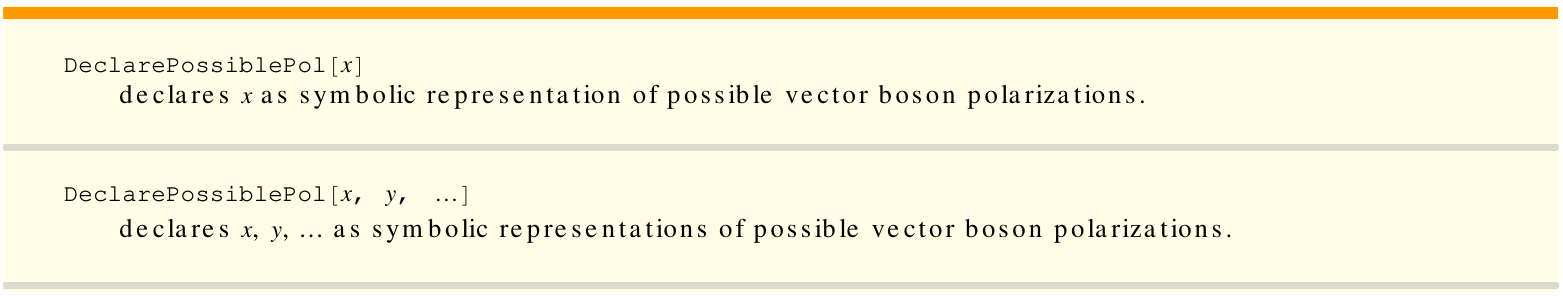}
\end{flushleft}

\subsubsection{UndeclarePossiblePol}

\label{SpinorsExtras/ref/UndeclarePossiblePol}

Removes given symbols from list of vector boson polarizations.

\begin{flushleft}
    \includegraphics{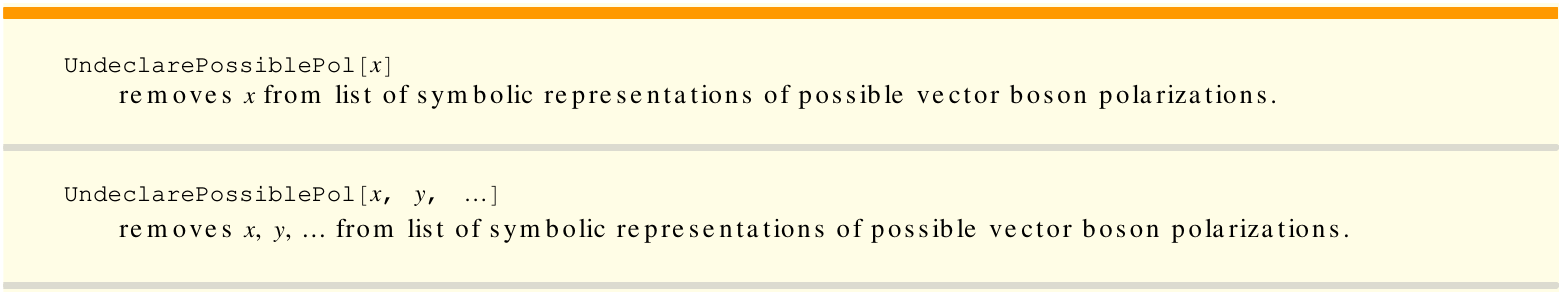}
\end{flushleft}

\subsubsection{PossiblePolQ}

\label{SpinorsExtras/ref/PossiblePolQ}

Tests whether given expression is interpretable as vector boson polarization.

\begin{flushleft}
    \includegraphics{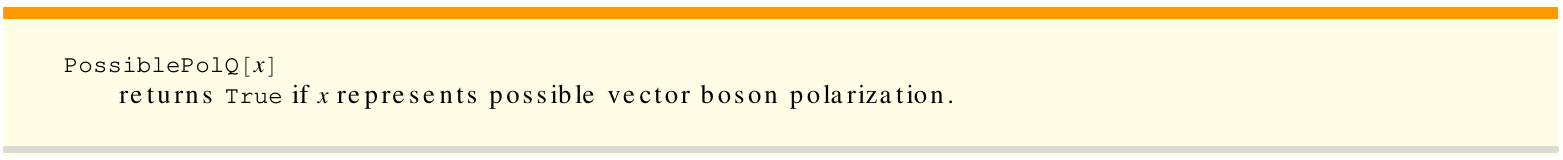}
\end{flushleft}

\subsection{Utilities}

\subsubsection{ReplaceLVector}

\label{SpinorsExtras/ref/ReplaceLVector}

Replaces given Lorentz vector in given expression.

\begin{flushleft}
    \includegraphics{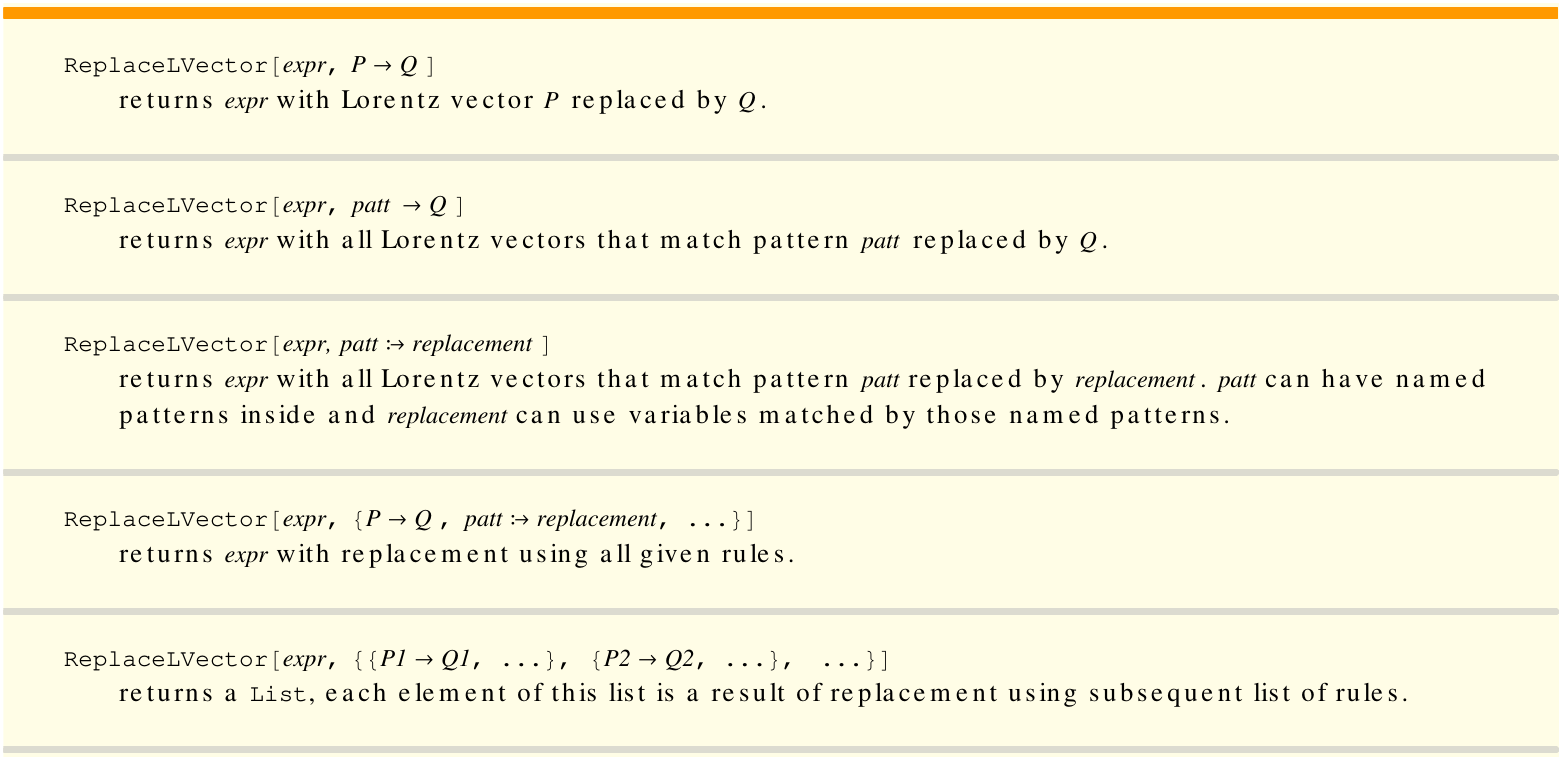}
    
    \includegraphics{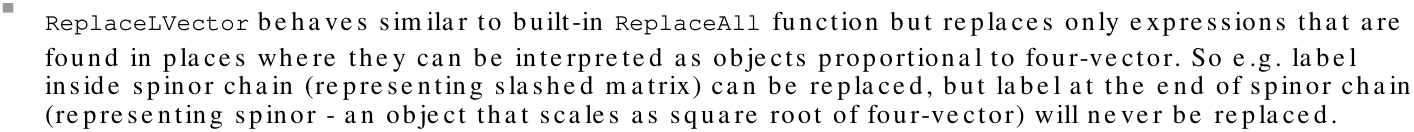}
\end{flushleft}

\subsubsection{ReplaceBSpinor}

\label{SpinorsExtras/ref/ReplaceBSpinor}

Replaces given massless or massive B spinor in given expression.

\begin{flushleft}
    \includegraphics{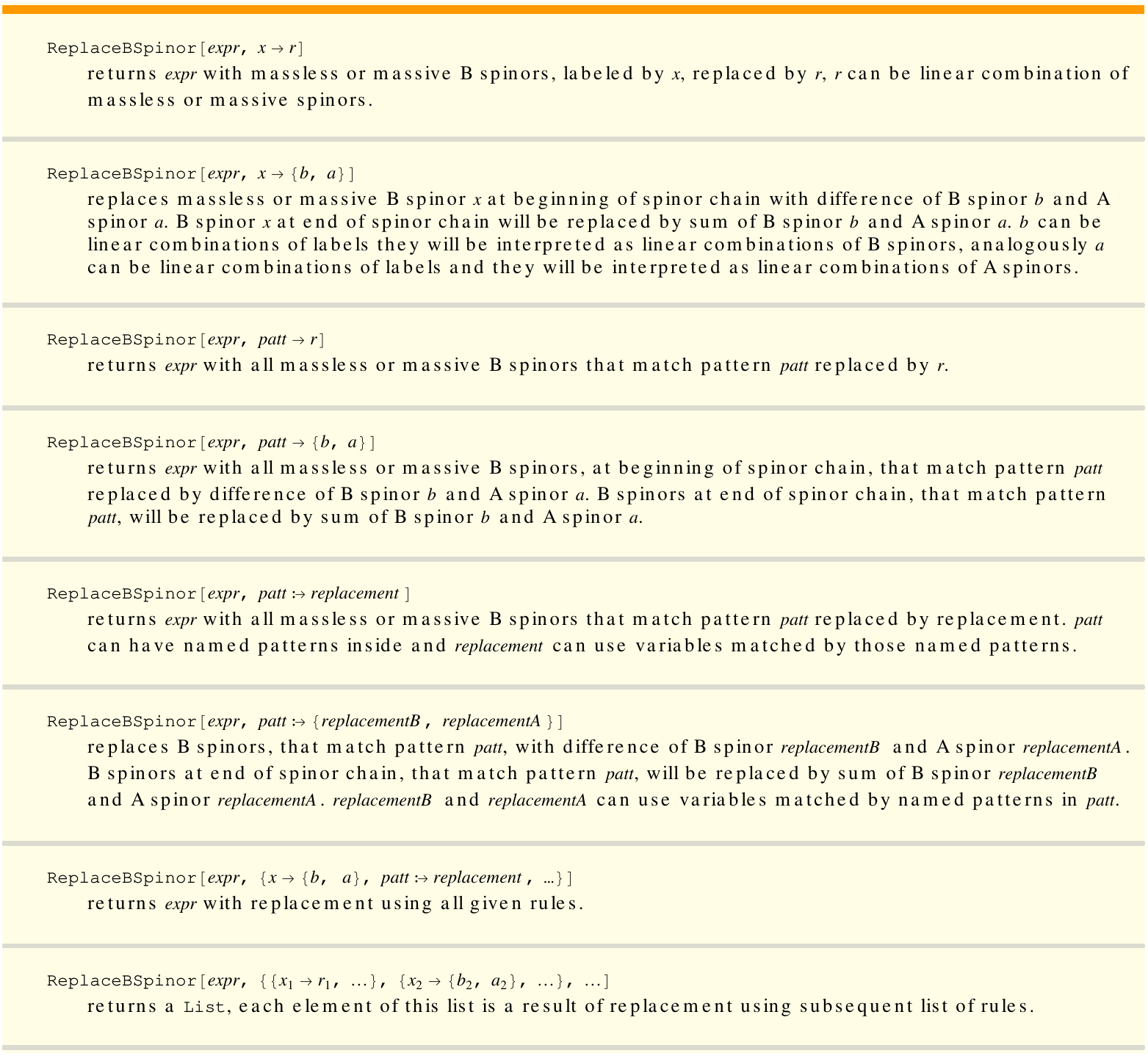}
    
    \includegraphics{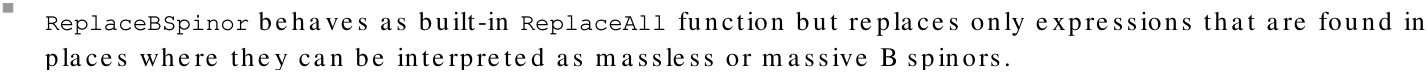}
\end{flushleft}

\subsubsection{ReplaceASpinor}

\label{SpinorsExtras/ref/ReplaceASpinor}

Replaces given massless or massive A spinor in given expression.

\begin{flushleft}
    \includegraphics{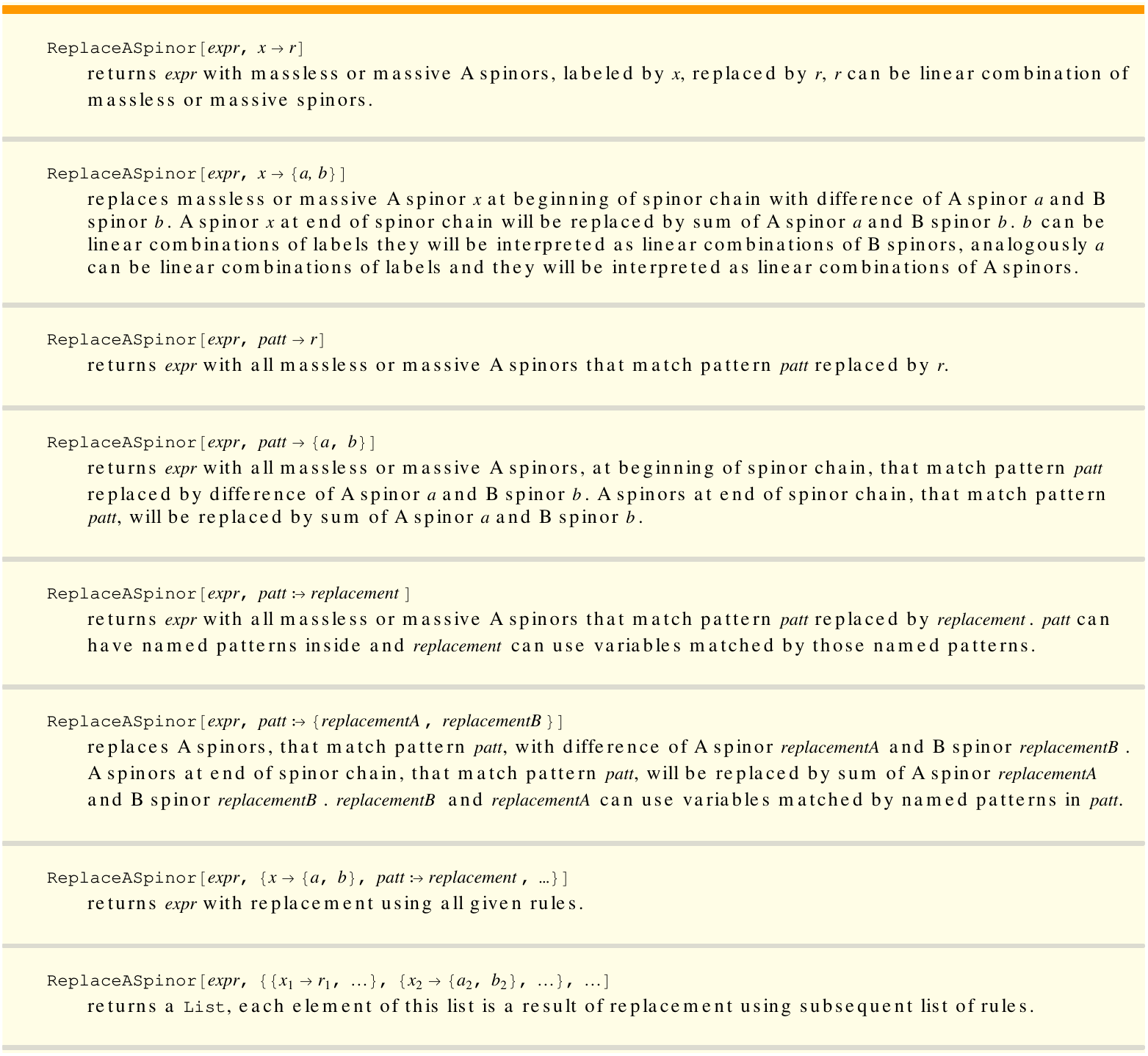}
    
    \includegraphics{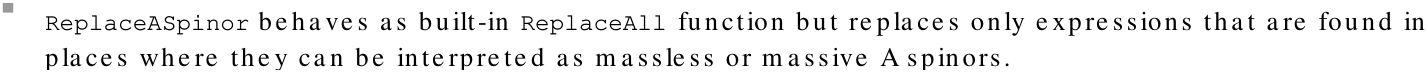}
\end{flushleft}

\subsubsection{ExpandMPToSpinors}

\label{SpinorsExtras/ref/ExpandMPToSpinors}

Replaces Minkowski products by spinor products.

\begin{flushleft}
    \includegraphics{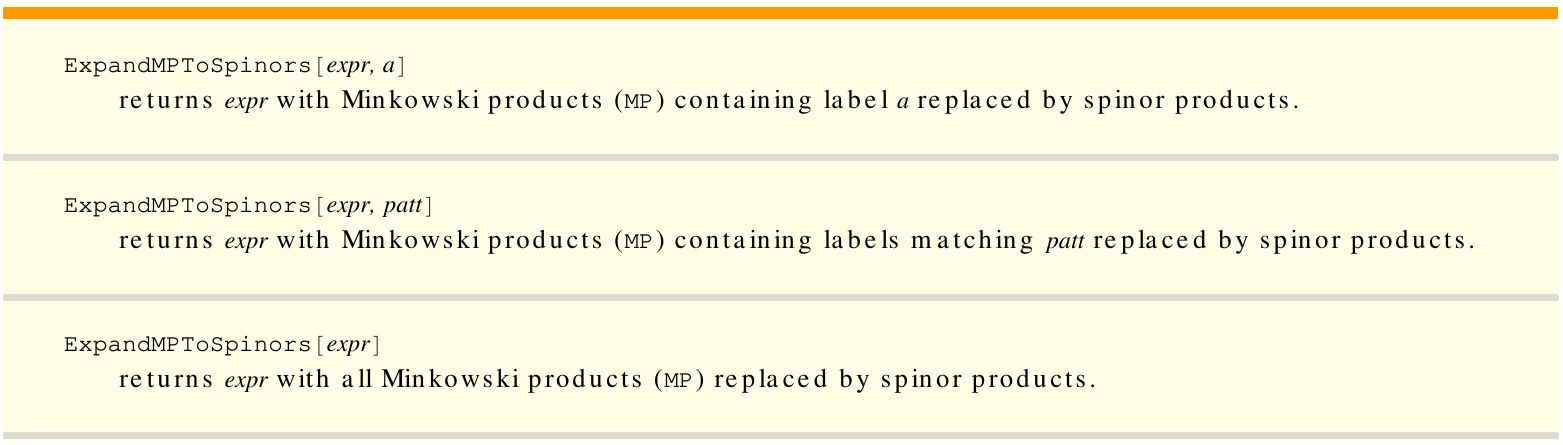}
    
    \includegraphics{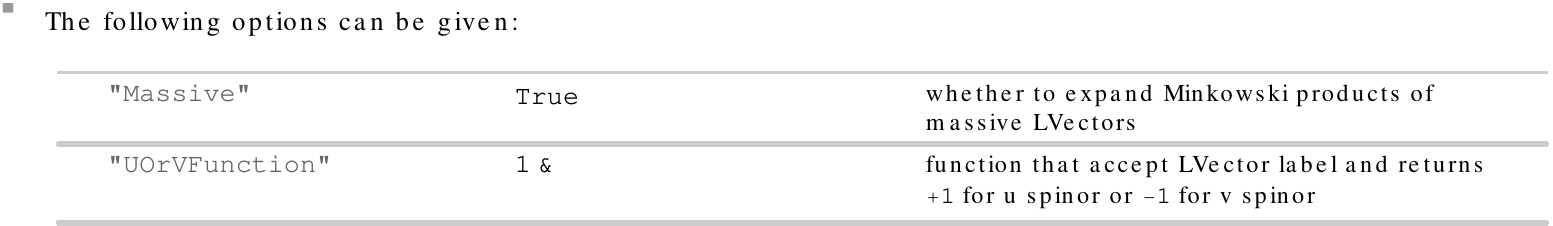}
\end{flushleft}

\subsubsection{ExpandSToMPs}

\label{SpinorsExtras/ref/ExpandSToMPs}

Replaces s invariants by Minkowski products.

\begin{flushleft}
    \includegraphics{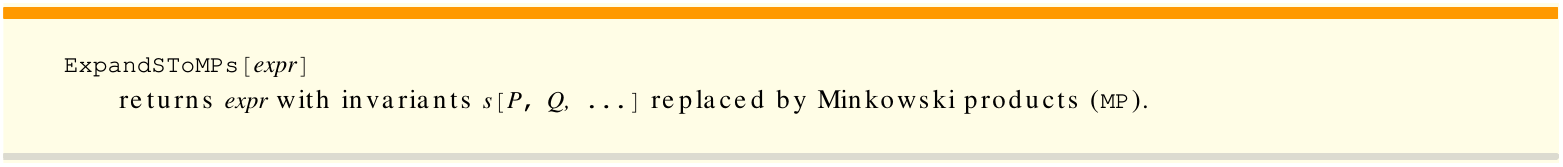}
    
    \includegraphics{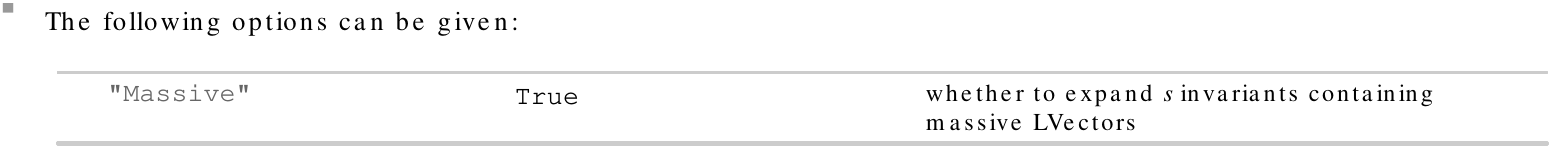}
\end{flushleft}

\subsubsection{DeclarePlusMinusOne}

\label{SpinorsExtras/ref/DeclarePlusMinusOne}

Sets given symbols to be treated as $\pm $1.

\begin{flushleft}
    \includegraphics{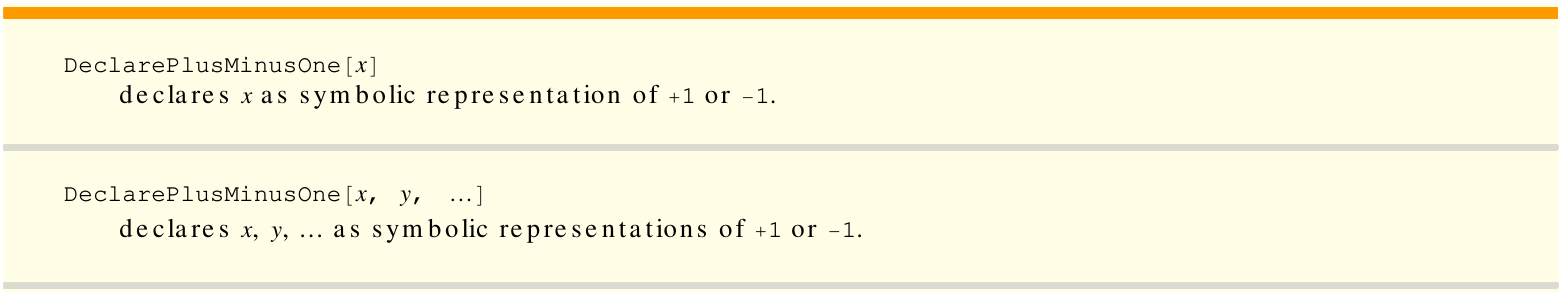}
\end{flushleft}

\subsubsection{UndeclarePlusMinusOne}

\label{SpinorsExtras/ref/UndeclarePlusMinusOne}

Removes given symbols from list of expressions treated as $\pm $1.

\begin{flushleft}
    \includegraphics{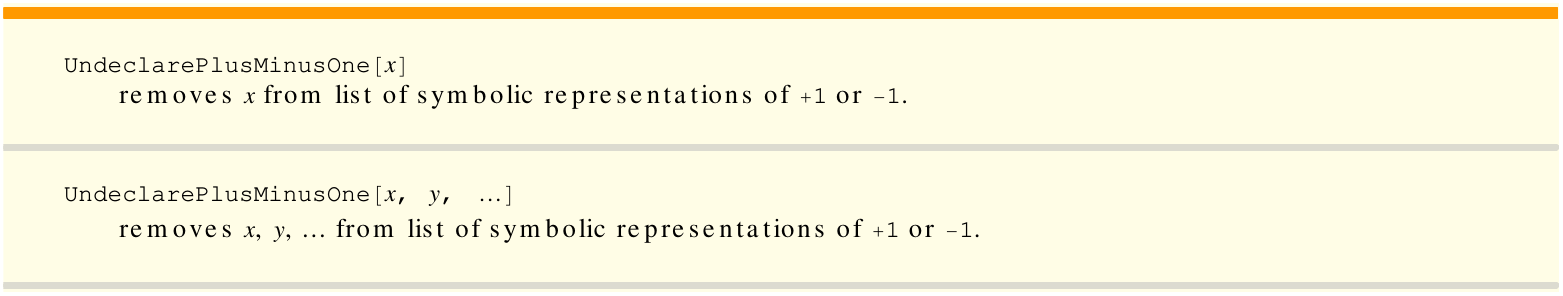}
\end{flushleft}

\subsubsection{PlusMinusOneQ}

\label{SpinorsExtras/ref/PlusMinusOneQ}

Tests whether given expression is interpretable as $\pm $1.

\begin{flushleft}
    \includegraphics{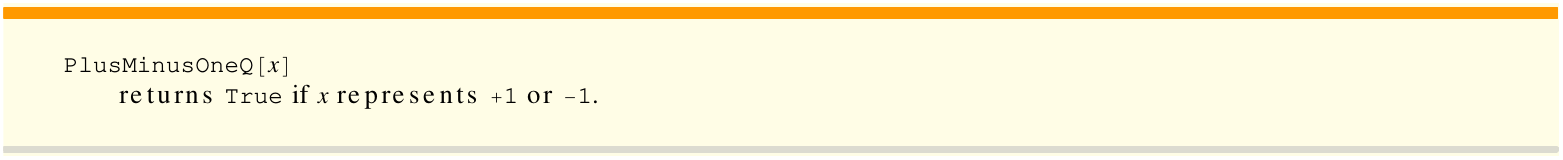}
    
    \includegraphics{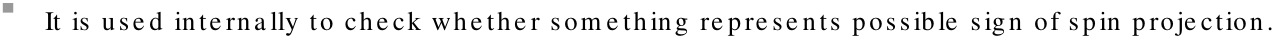}
\end{flushleft}

\subsubsection{AnySpinorQ}

\label{SpinorsExtras/ref/AnySpinorQ}

Tests whether given expression is interpretable as massless or massive spinor.

\begin{flushleft}
    \includegraphics{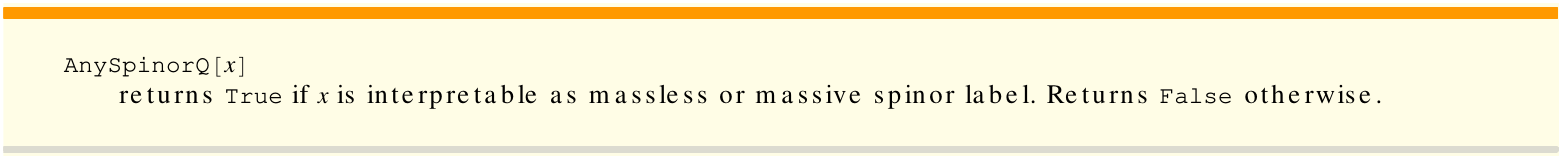}
\end{flushleft}

\subsection{Proportional spinors and vectors}

\subsubsection{DeclareBSpinorProportional}

\label{SpinorsExtras/ref/DeclareBSpinorProportional}

Declares that B spinors with given labels are proportional.

\begin{flushleft}
    \includegraphics{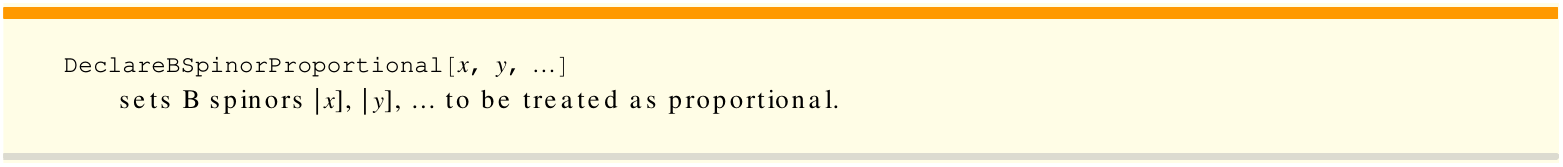}
\end{flushleft}

\subsubsection{DeclareASpinorProportional}

\label{SpinorsExtras/ref/DeclareASpinorProportional}

Declares that A spinors with given labels are proportional.

\begin{flushleft}
    \includegraphics{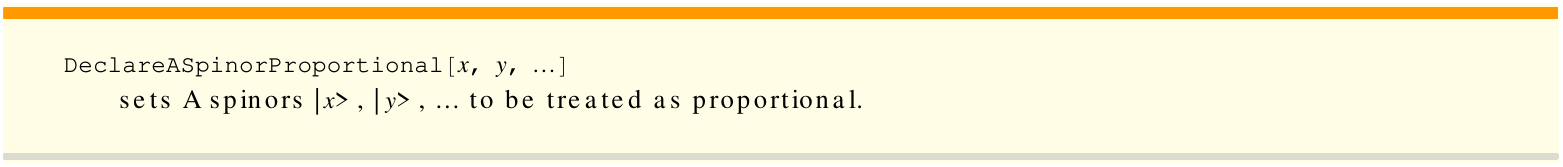}
\end{flushleft}

\subsubsection{DeclareLVectorProportional}

\label{SpinorsExtras/ref/DeclareLVectorProportional}

Declares that LVectors with given labels are proportional.

\begin{flushleft}
    \includegraphics{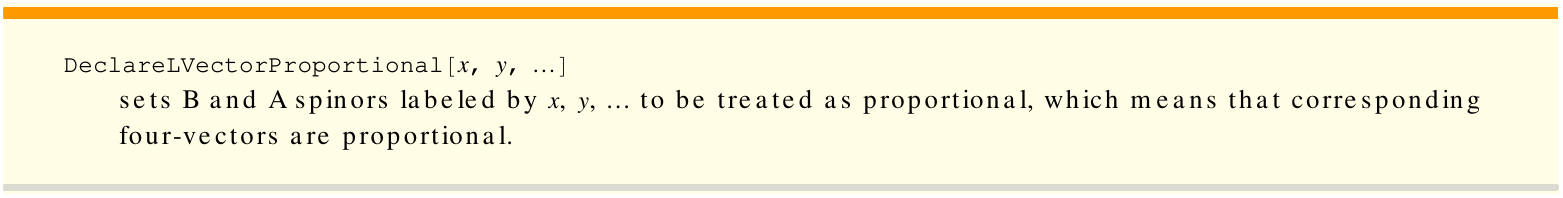}
\end{flushleft}

\subsubsection{BSpinorProportionalQ}

\label{SpinorsExtras/ref/BSpinorProportionalQ}

Tests whether B spinors with given labels are proportional.

\begin{flushleft}
    \includegraphics{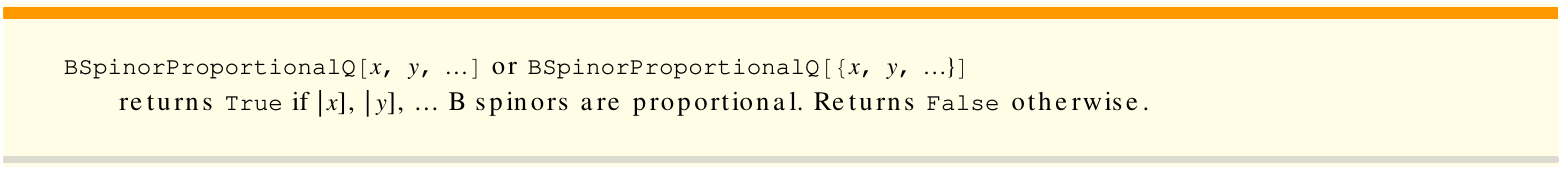}
\end{flushleft}

\subsubsection{ASpinorProportionalQ}

\label{SpinorsExtras/ref/ASpinorProportionalQ}

Tests whether A spinors with given labels are proportional.

\begin{flushleft}
    \includegraphics{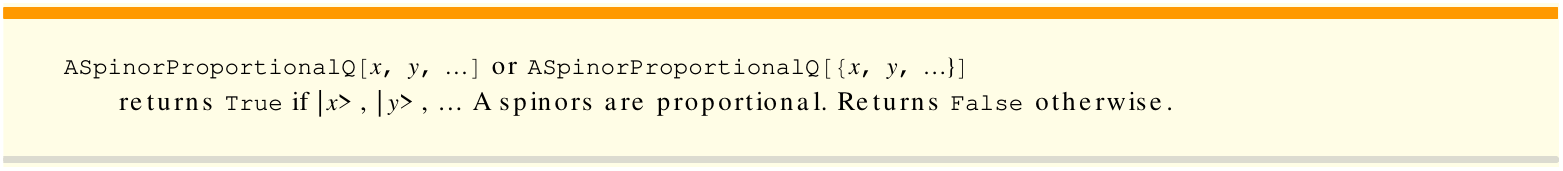}
\end{flushleft}

\subsubsection{LVectorProportionalQ}

\label{SpinorsExtras/ref/LVectorProportionalQ}

Tests whether LVectors with given labels are proportional.

\begin{flushleft}
    \includegraphics{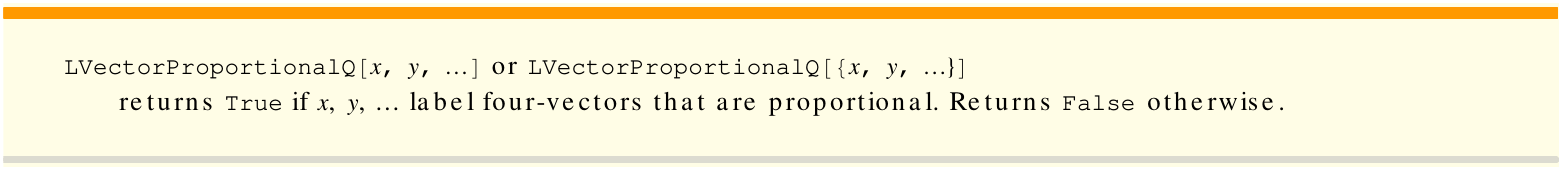}
    
    \includegraphics{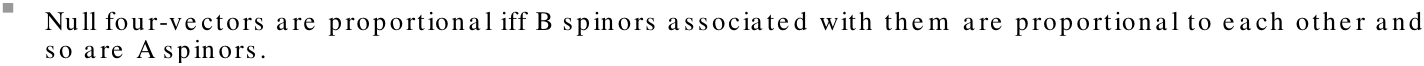}
\end{flushleft}

\subsection{Phases management}

\subsubsection{AppendPhase}

\label{SpinorsExtras/ref/AppendPhase}

Multiplies parts of expression with additional phases.

\begin{flushleft}
    \includegraphics{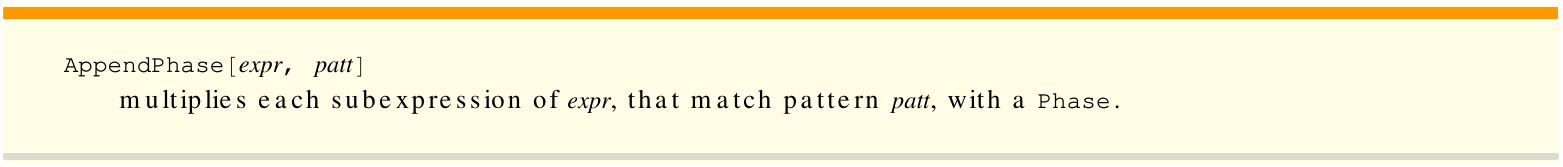}
    
    \includegraphics{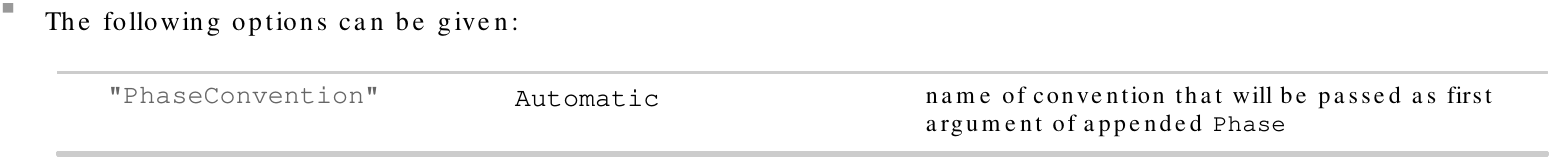}
\end{flushleft}

\subsubsection{Phase}

\label{SpinorsExtras/ref/Phase}

Represents additional phase of given expression.

\begin{flushleft}
    \includegraphics{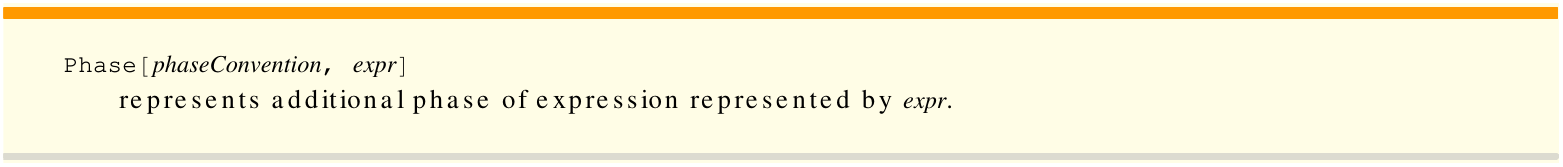}
\end{flushleft}

\subsection{Spinor decomposition}

\subsubsection{DecomposeBSpinor}

\label{SpinorsExtras/ref/DecomposeBSpinor}

Decomposes B spinor in given basis.

\begin{flushleft}
    \includegraphics{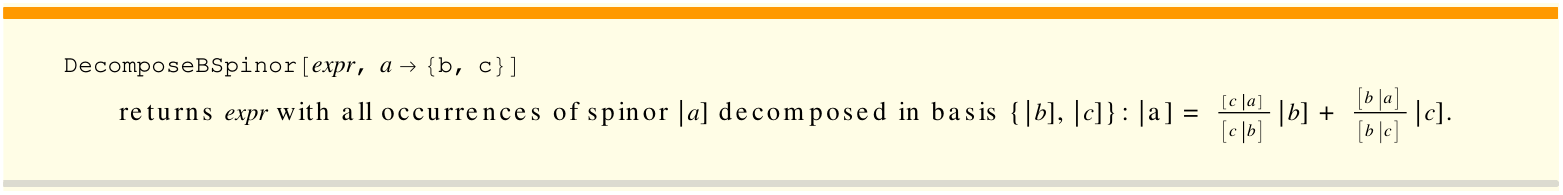}
\end{flushleft}

\subsubsection{DecomposeASpinor}

\label{SpinorsExtras/ref/DecomposeASpinor}

Decomposes A spinor in given basis.

\begin{flushleft}
    \includegraphics{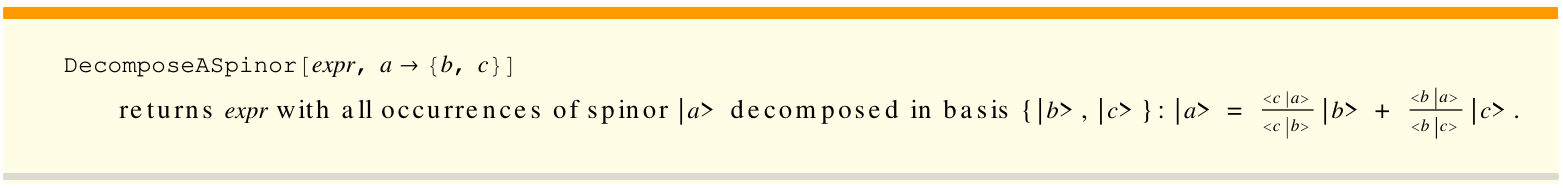}
\end{flushleft}

\subsection{Simple Tensors}

\subsubsection{SimpleTensorQ}

\label{SpinorsExtras/ref/SimpleTensorQ}

Tests whether given expression represents simple tensor.

\begin{flushleft}
    \includegraphics{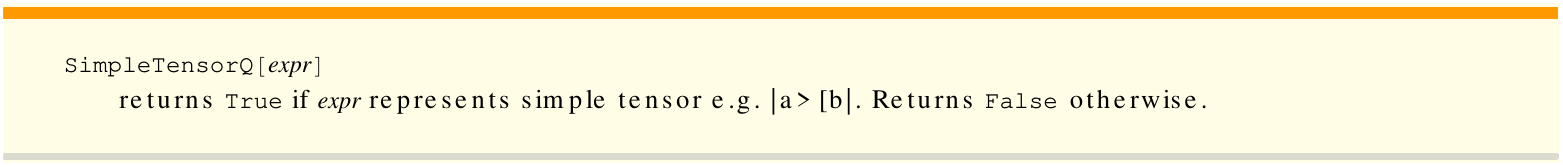}
\end{flushleft}

\subsubsection{SimpleTensorGetBLabel}

\label{SpinorsExtras/ref/SimpleTensorGetBLabel}

Extracts B spinor from tensor product of B and A spinors.

\begin{flushleft}
    \includegraphics{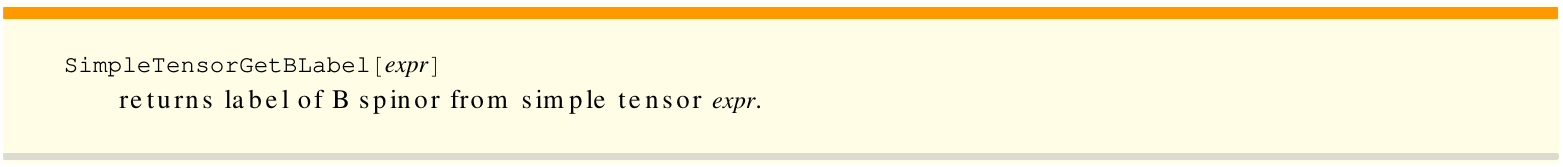}
\end{flushleft}

\subsubsection{SimpleTensorGetALabel}

\label{SpinorsExtras/ref/SimpleTensorGetALabel}

Extracts A spinor from tensor product of B and A spinors.

\begin{flushleft}
    \includegraphics{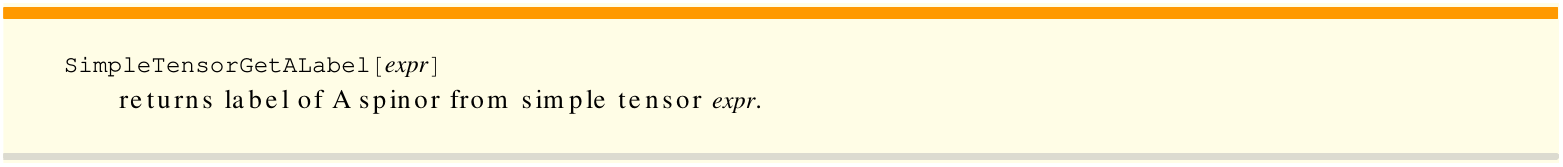}
    
    \includegraphics{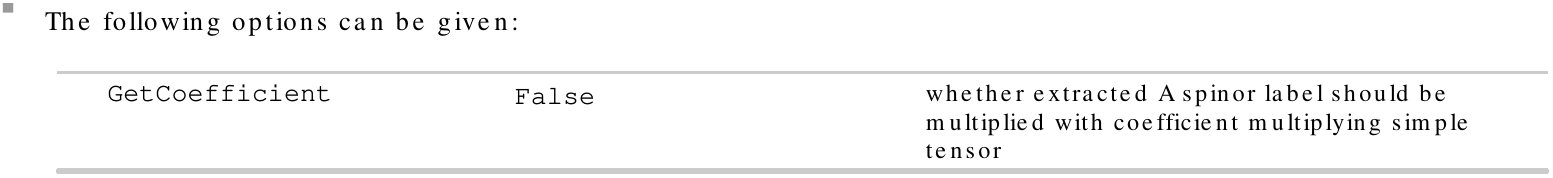}
\end{flushleft}

\subsection{Numerics}

\subsubsection{DeclareSpinorRandomMomentum}

\label{SpinorsExtras/ref/DeclareSpinorRandomMomentum}

Generates random numerics for given spinor.

\begin{flushleft}
    \includegraphics{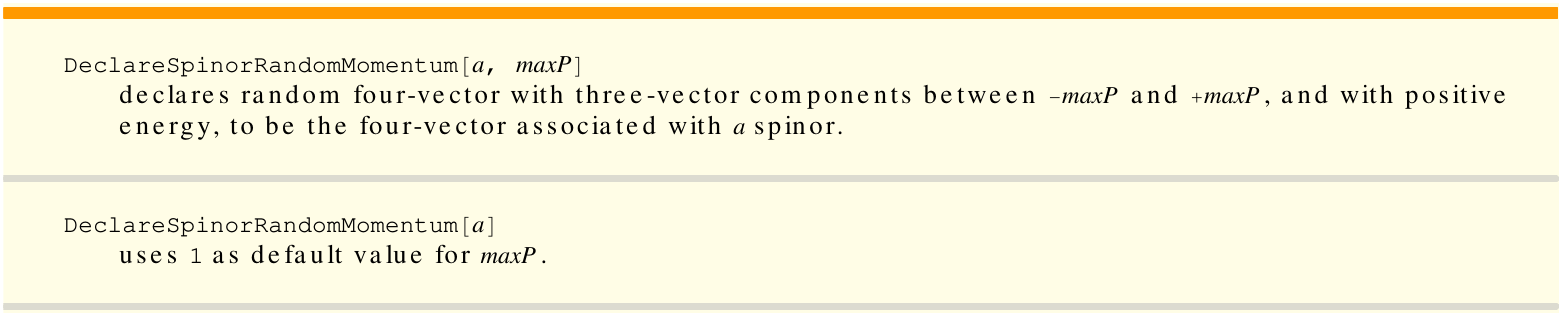}
    
    \includegraphics{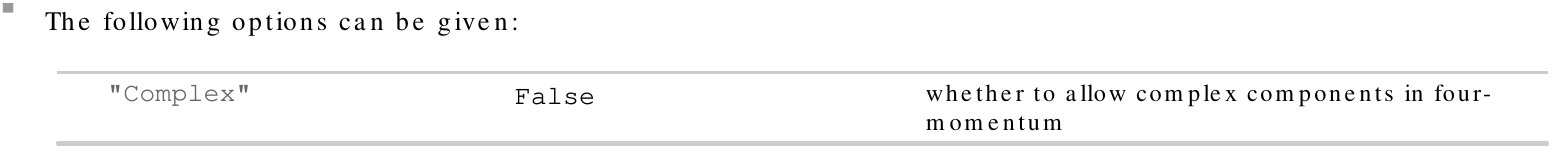}
\end{flushleft}

\subsubsection{GenComplexMomenta}

\label{SpinorsExtras/ref/GenComplexMomenta}

Generates random complex momenta for spinors so that they sum to zero.

\begin{flushleft}
    \includegraphics{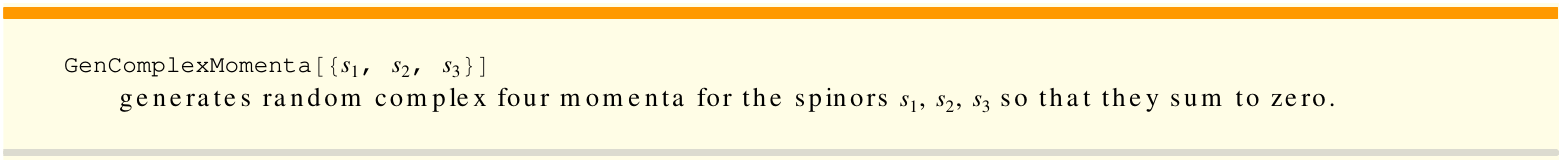}
    
    \includegraphics{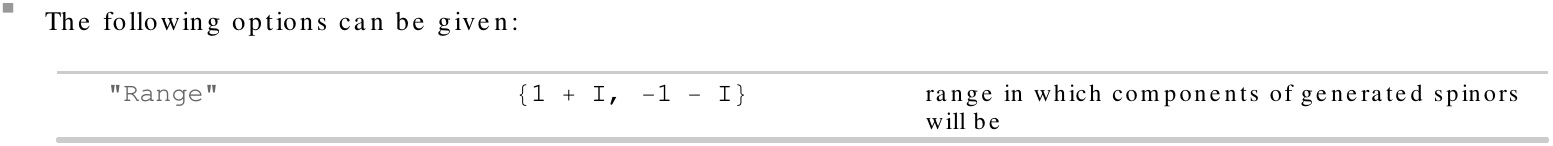}
\end{flushleft}

\subsection{Functions from original Spinors\`{ } context with modified behavior}

\subsubsection{SpOpen}

\label{SpinorsExtras/ref/SpOpen}

Decomposes spinor chains to products of smaller spinor chains.

\begin{flushleft}
    \includegraphics{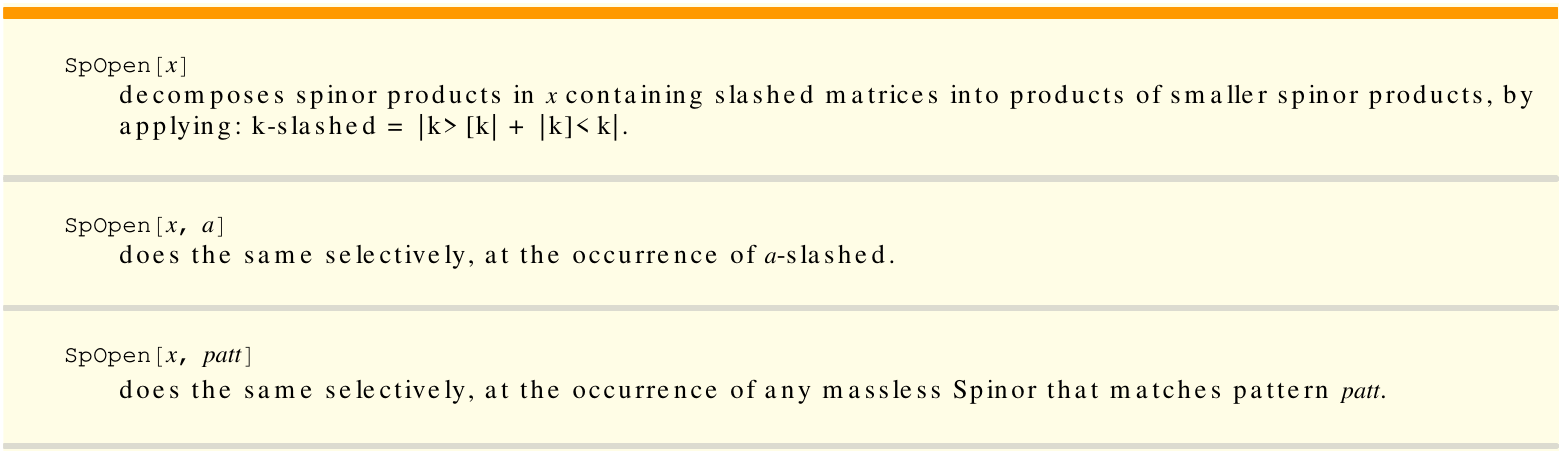}
    
    \includegraphics{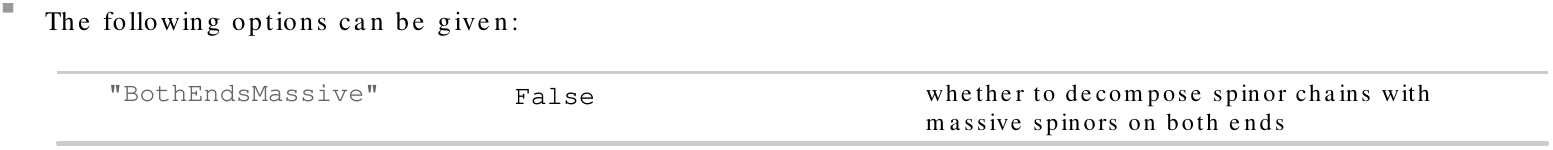}
\end{flushleft}

\subsubsection{ExpandSToSpinors}

\label{SpinorsExtras/ref/ExpandSToSpinors}

Converts s invariants to products of spinor chains.

\begin{flushleft}
    \includegraphics{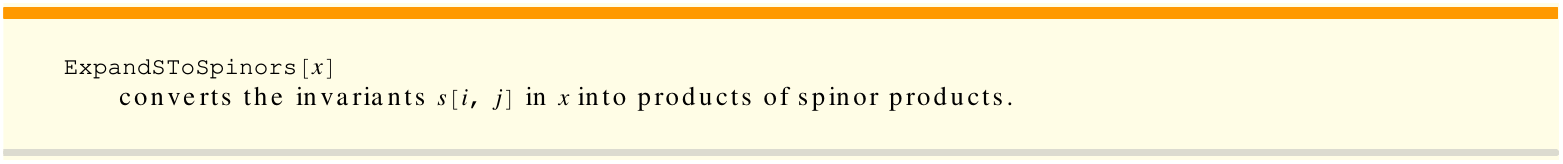}
    
    \includegraphics{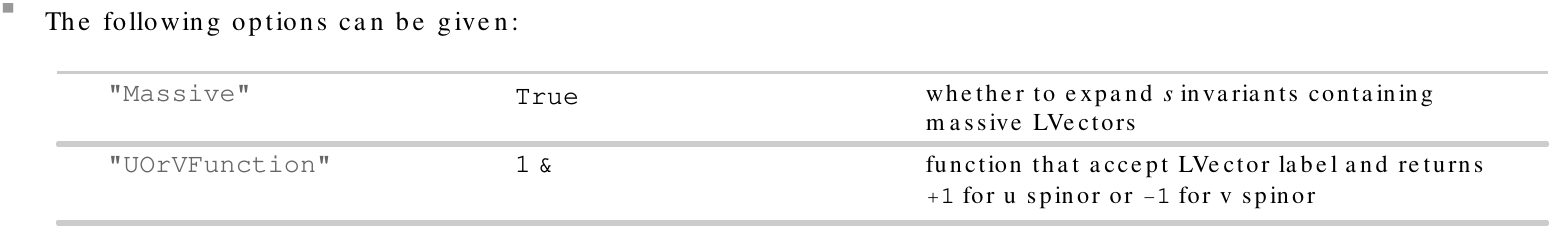}
\end{flushleft}

\section{Tutorial}
\label{sec:tutorial}

In this Section we illustrate how \texttt{SpinorsExtras} package can be
used for calculation of amplitudes including massive spinors and
vector bosons on the simple case of QED with photons, electrons and
muons.  As an example, we calculate the tree level amplitude
for \(e^-\mu ^-\to e^-\mu ^-\) scattering using on-shell recursion
method. Than we compare it with amplitude calculated using Feynman
diagrams and test invariance with respect to change of various
reference vectors.  The same example is included in the form of
Mathematica tutorial in \texttt{SpinorsExtras} package, and after installing
it, can be found in Mathematica Documentation Center e.g. by searching for
``SpinorsExtras tutorial''.

Throughout this tutorial basic familiarity with S@M package is assumed and only functions added in SpinorsExtras package are introduced in definition
boxes.

To use SpinorsExtras, one needs to import the package first:

\begin{flushleft}
    \includegraphics{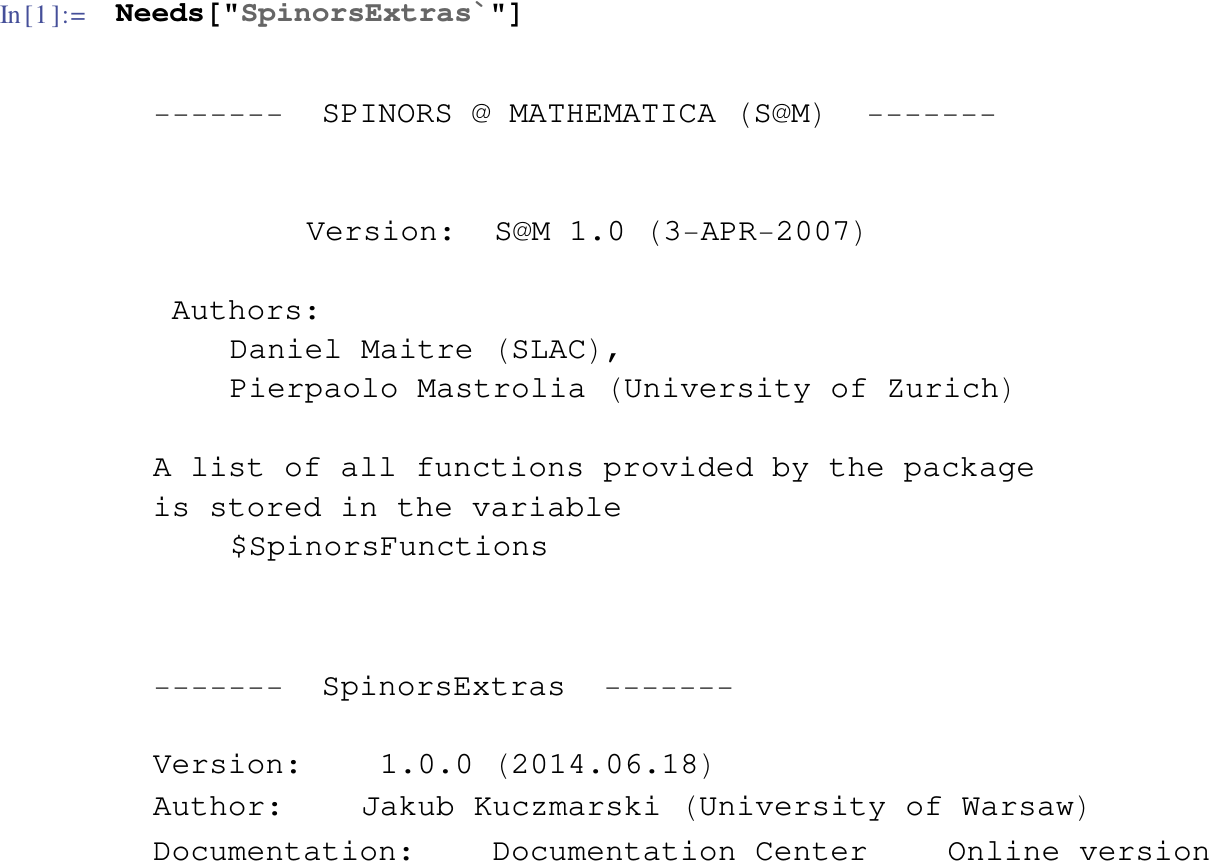}
\end{flushleft}

\subsection{On-shell recursion}
\setcounter{tutorialStep}{0}

\begin{table}
    \includegraphics{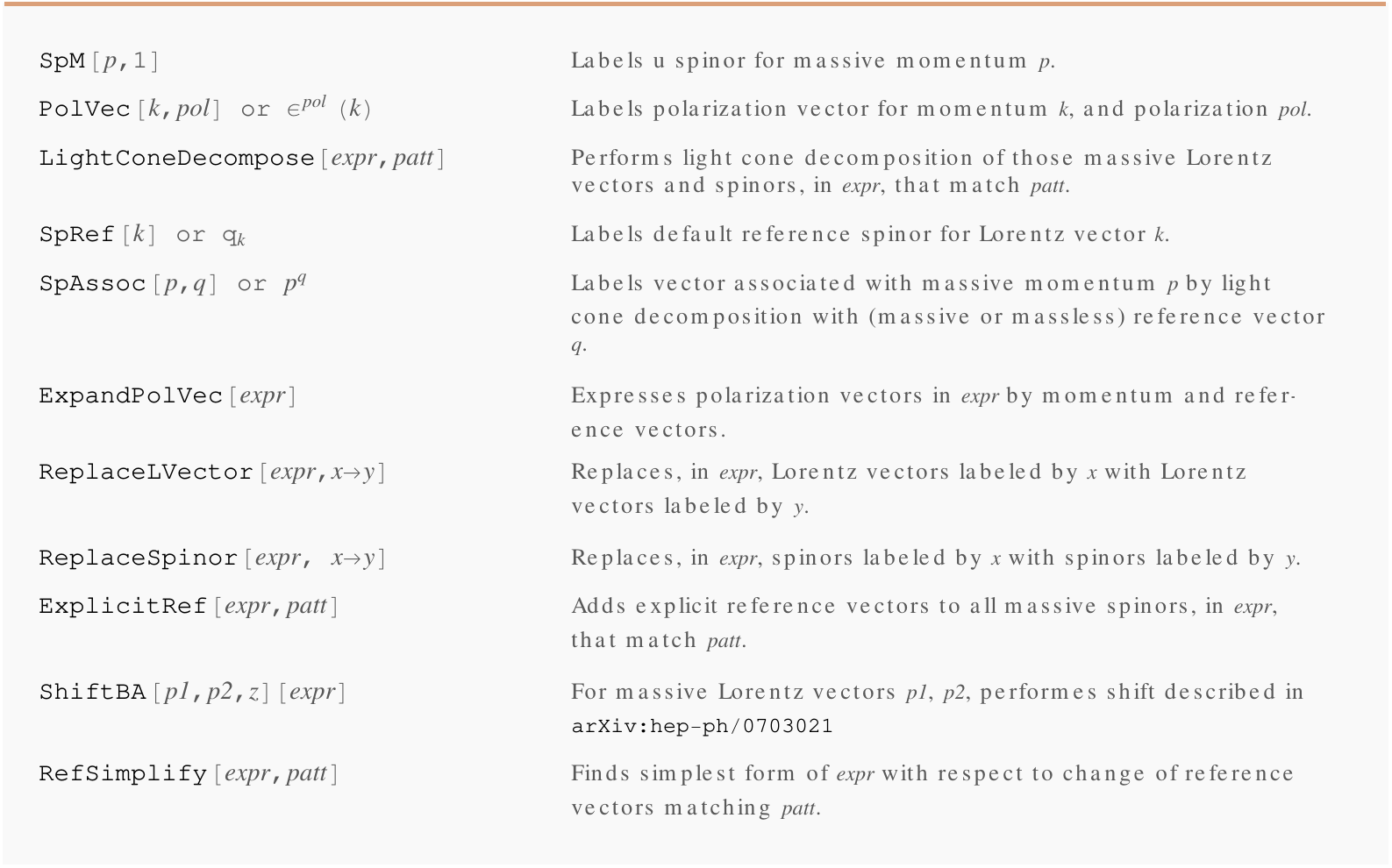}
    \caption{Functions used to calculate amplitudes using on-shell recursion.}
\end{table}

We start from performing the amplitude calculation using the on-shell recursion technique. It can be done with following steps:

\tutorialStep Define basic fermion-vector boson-fermion three point amplitudes:

\begin{flushleft}
    \includegraphics{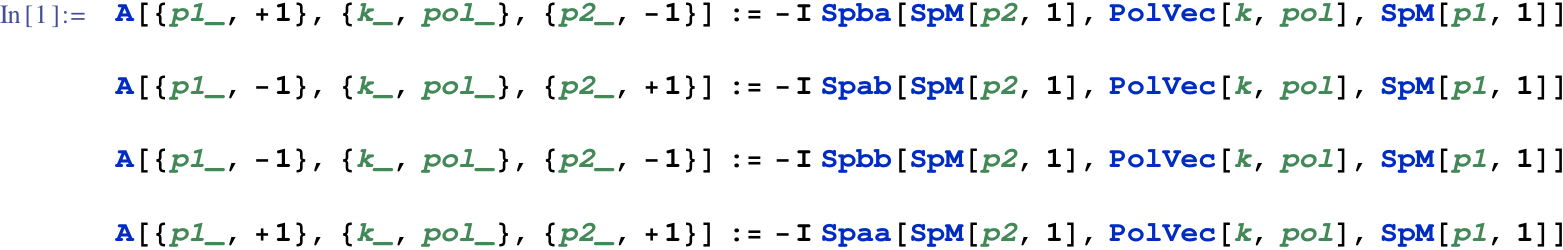}
\end{flushleft}

\tutorialStep Declare symbol to be treated as massless labels (\lstinline!k! and \lstinline!mk! will denote momentum of on-shell photon):

\begin{flushleft}
    \includegraphics{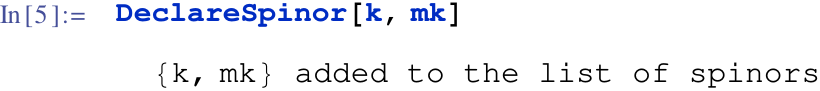}
\end{flushleft}

\tutorialStep Declare symbols to be treated as massive labels (\lstinline!p1!, \lstinline!p3! will denote on-shell momenta of massive electrons;
\lstinline!p2!, \lstinline!p4! momenta of muons; \lstinline!kOff! will denote momentum of off-shell photon):

\begin{flushleft}
    \includegraphics{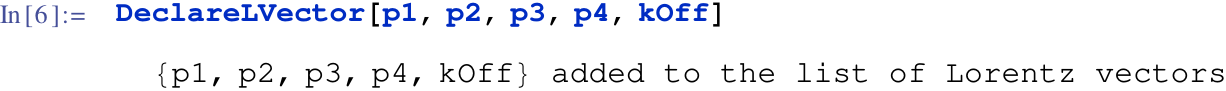}
\end{flushleft}

\tutorialStep Assign masses:

\begin{flushleft}
    \includegraphics{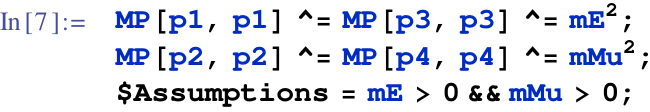}
\end{flushleft}

\tutorialStep Calculate electron muon scattering diagrams with on-shell photon \(\text{\textit{$t$}}\)-channel exchange.

\begin{flushleft}
    \includegraphics{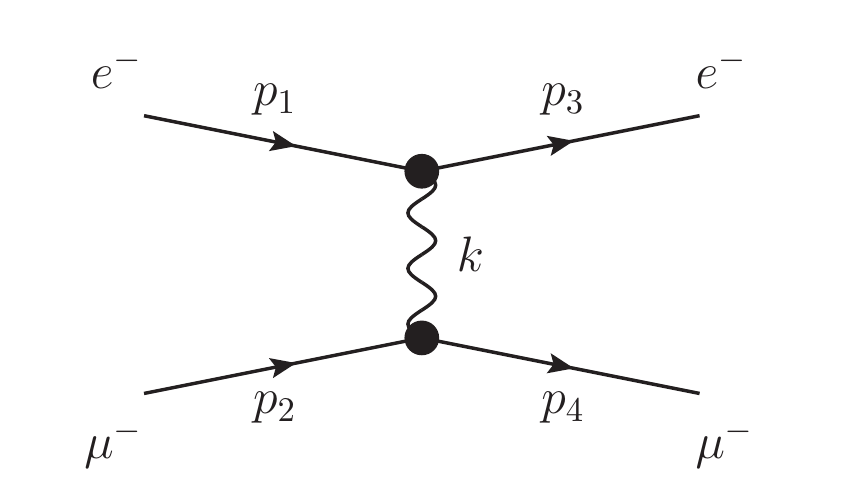}
\end{flushleft}

Electrons have positive and muons negative spin projections. Diagram consists of two fermion-vector boson-fermion three point amplitudes connected
by photon line. Momentum transfer is denoted by \lstinline!k!.

Symbols from S@M package are linear in labels. We a priori don't know whether label representing four-momentum \lstinline!-k! will appear in place
where it'll be interpreted as object scaling linearly with four-momentum, or as object scaling as square root of four-momentum. In first case it
should be labeled by \lstinline!-k! and in latter case by $\pm $\lstinline!I k!, so we temporarily use \lstinline!mk! to represent it.

\begin{flushleft}
    \includegraphics{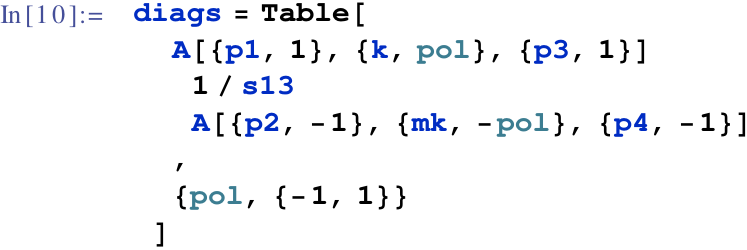}
\end{flushleft}

\begin{flushleft}
    \includegraphics{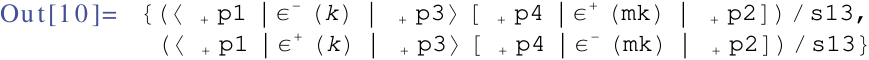}
\end{flushleft}

\tutorialStep Decompose spinors for {``}final{''} particles:

\begin{flushleft}
    \includegraphics{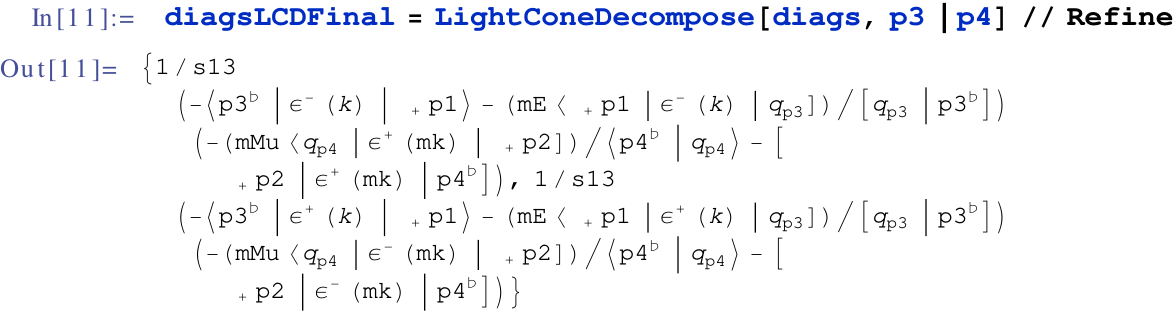}
\end{flushleft}

\tutorialStep Express polarization vectors by momentum and reference vectors:

\begin{flushleft}
    \includegraphics{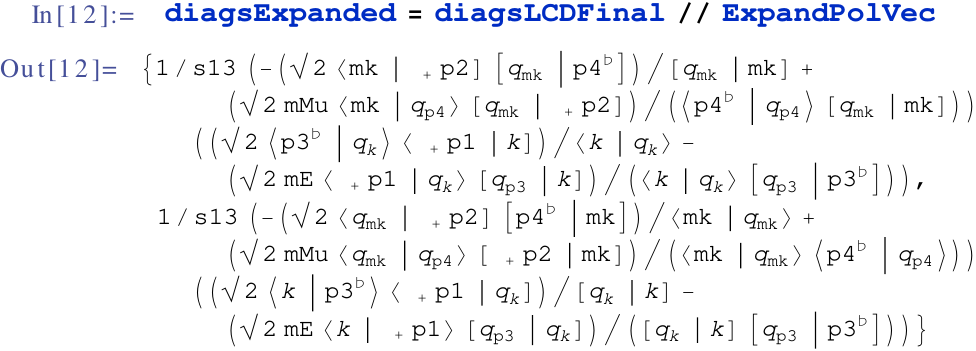}
\end{flushleft}

Lorentz vectors labeled by \(mk\) can now be replaced with \(-k\) and spinors labeled by \(mk\) can be replaced with $\pm $\(I k\) (in this simple
example there are actually no Lorentz vectors \(mk\) only spinors):

\begin{flushleft}
    \includegraphics{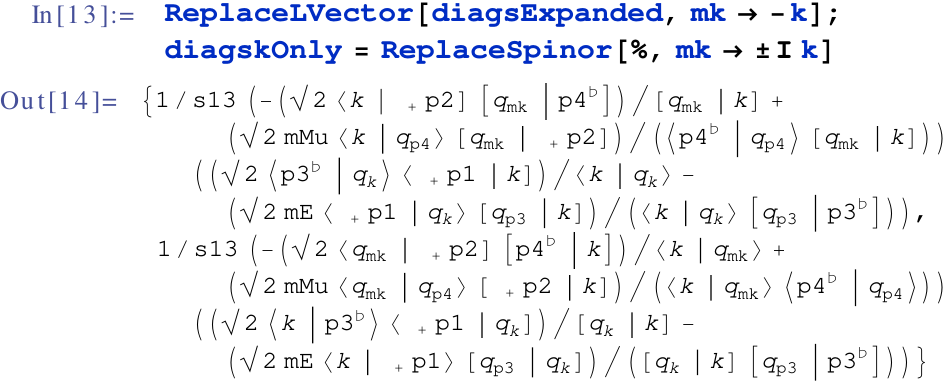}
\end{flushleft}

\tutorialStep Add explicit reference vectors to massive spinors that we intend to shift:

\begin{flushleft}
    \includegraphics{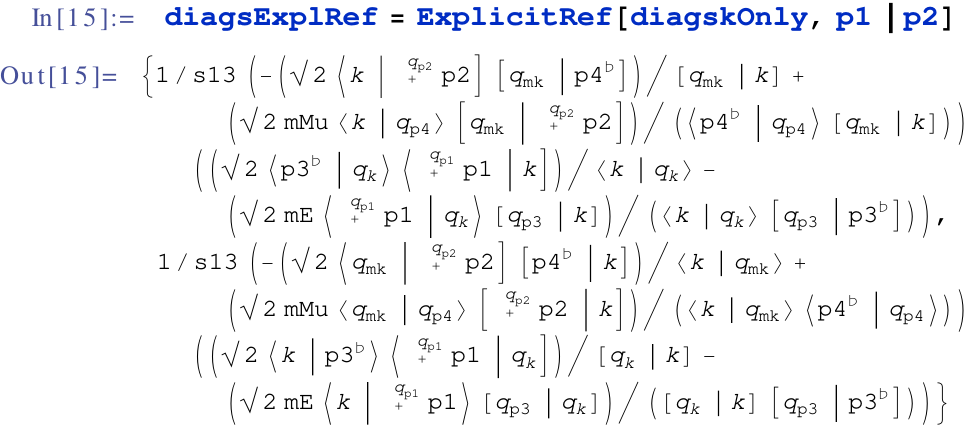}
\end{flushleft}

\tutorialStep Replace default reference vectors with reference vectors for which boundary term, for further shift, will vanish:

\begin{flushleft}
    \includegraphics{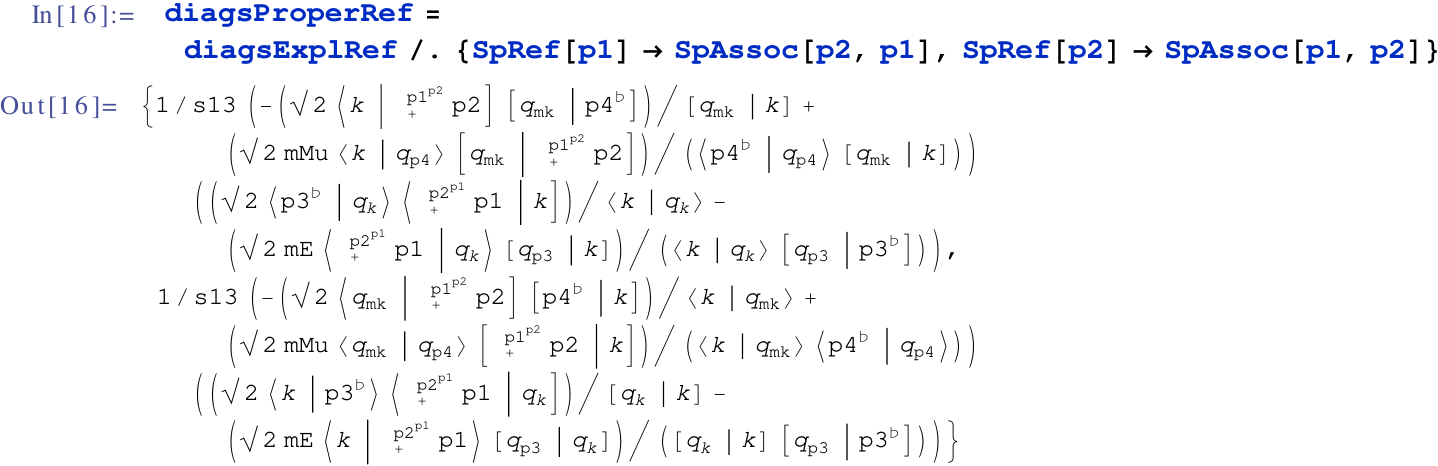}
\end{flushleft}

\tutorialStep Find \lstinline!z! that will put shifted momentum transfer on-shell:

\begin{flushleft}
    \includegraphics{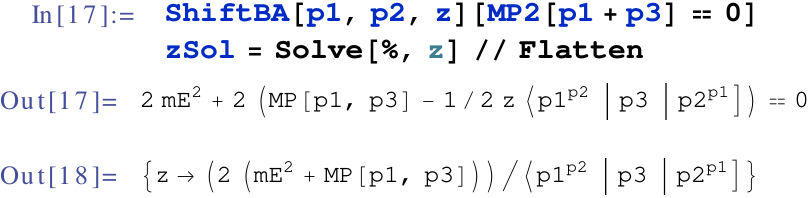}
\end{flushleft}

\tutorialStep Perform holomorphic shift of \lstinline!p1! and \lstinline!p2! in diagrams (in this simplified example there is actually nothing to
shift except momentum transfer which we'll consider separately) and insert proper \lstinline!z!:

\begin{flushleft}
    \includegraphics{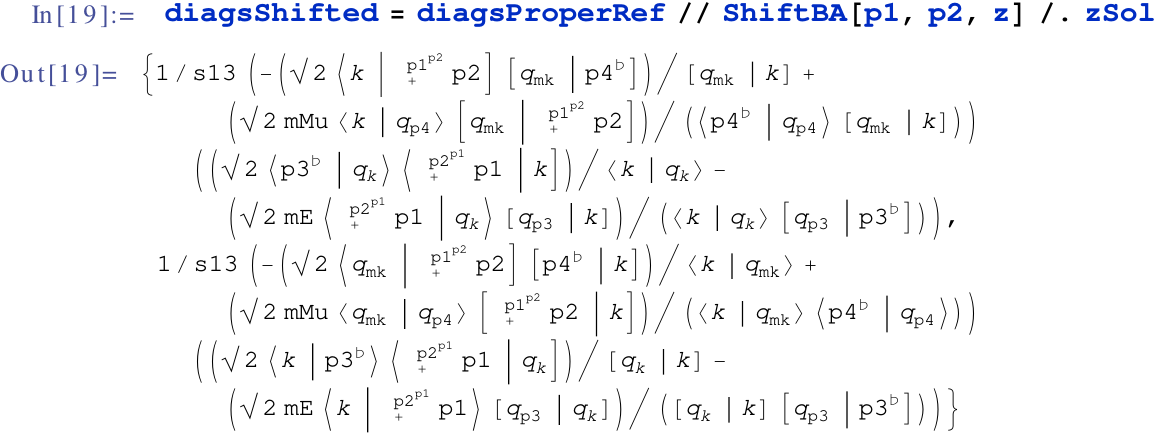}
\end{flushleft}

\tutorialStep After shift we can decompose remaining massive spinors:

\begin{flushleft}
    \includegraphics{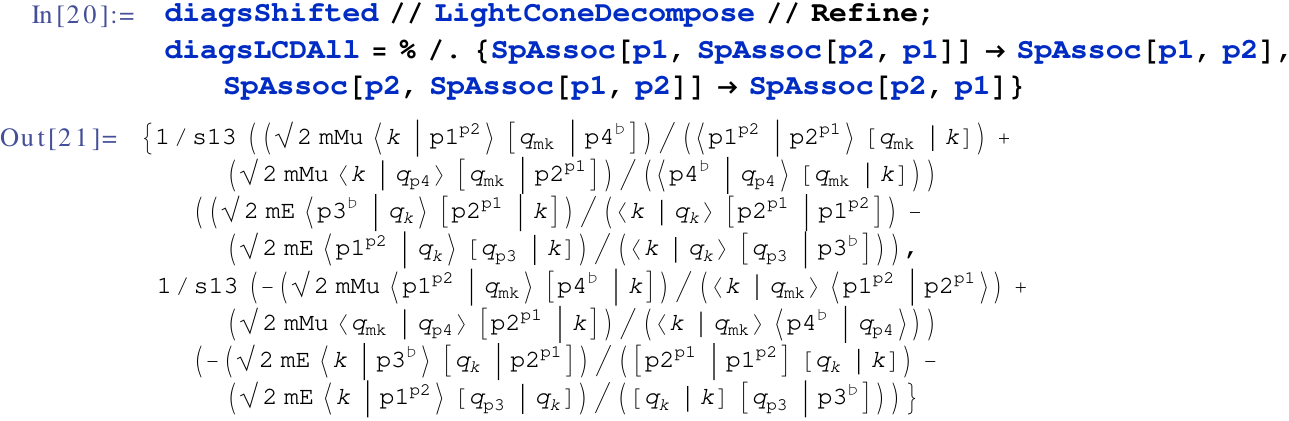}
\end{flushleft}

Spinors related to shifted, on-shell momentum transfer are proportional to slashed off-shell momentum transfer acting on proper spinors related to
complexification vector (normalization factor can be added to any spinor):

\begin{flushleft}
    \includegraphics{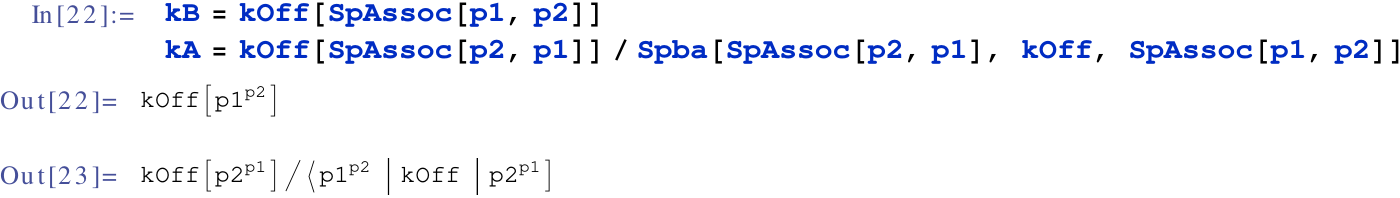}
\end{flushleft}

\tutorialStep Express photons momentum by external momenta:

\begin{flushleft}
    \includegraphics{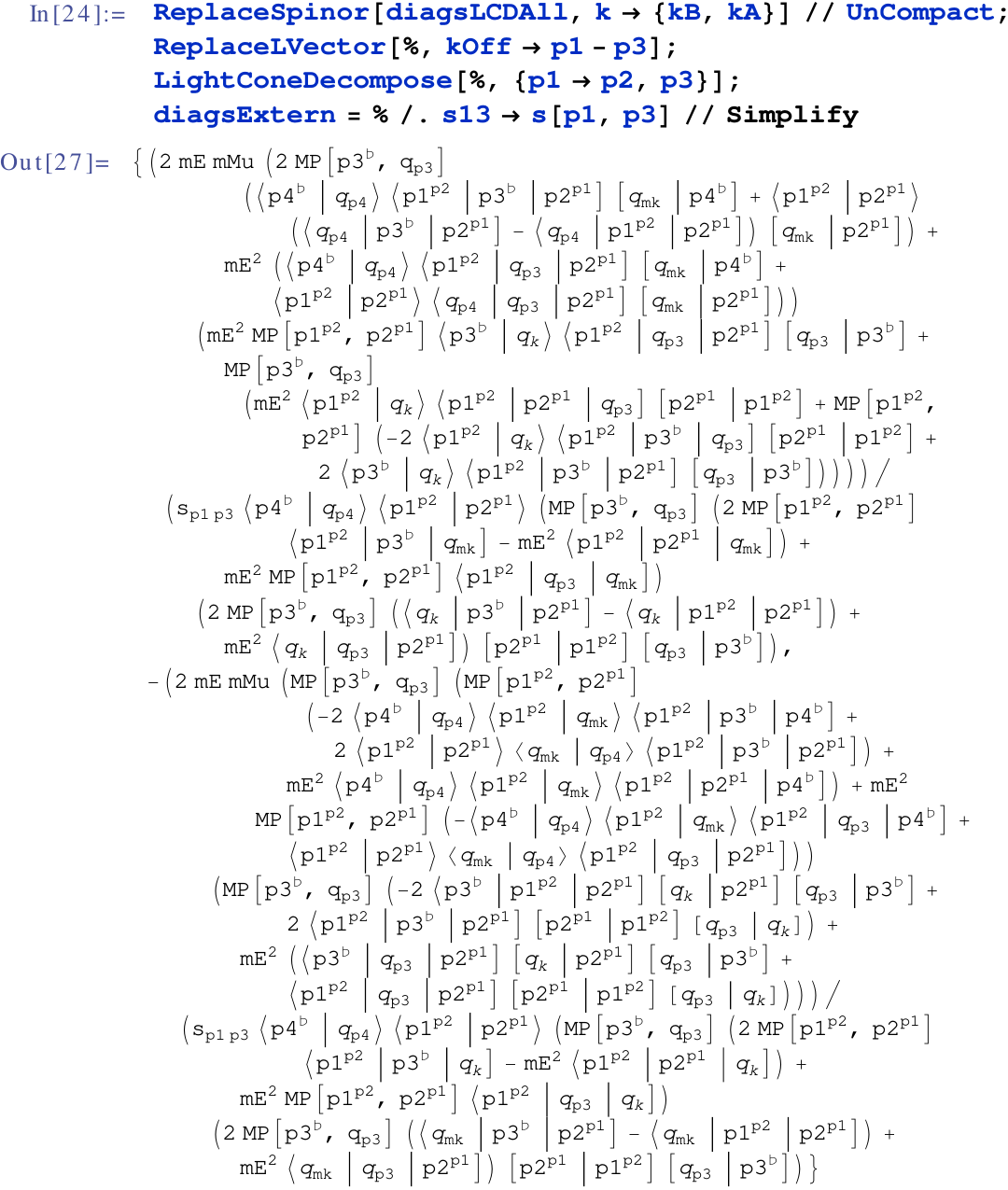}
\end{flushleft}

\tutorialStep Find simplest form of diagrams with respect to reference vectors of exchanged photon (this may take some time).

Note that since each diagram is gauge invariant we can simplify diagrams separately, so we map \hyperref[SpinorsExtras/ref/RefSimplify]{RefSimplify}
on list of diagrams:

\begin{flushleft}
    \includegraphics{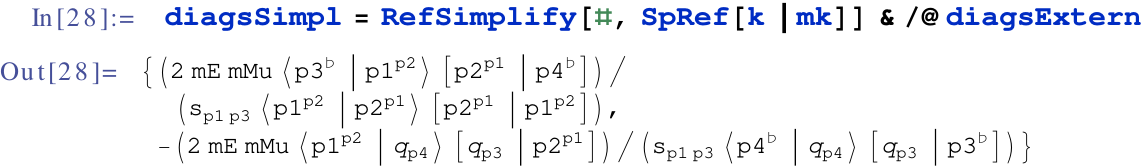}
\end{flushleft}

\tutorialStep Sum diagrams :

\begin{flushleft}
    \includegraphics{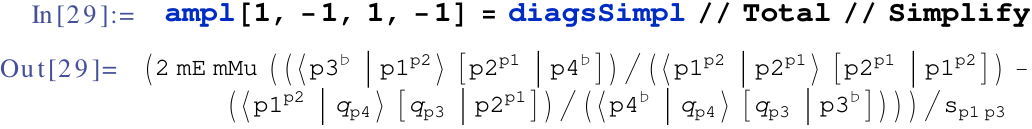}
\end{flushleft}

\subsection{Numerical results and comparison with standard amplitude}
\setcounter{tutorialStep}{0}

As we illustrate below SpinorsExtras package can be also used to calculate numerical values of scattering amplitudes. They can be used e.g. for testing
the final result by comparing it with amplitude obtained using ordinary Feynman diagrams.

\tutorialStep Assign numerical values to masses:

\begin{flushleft}
    \includegraphics{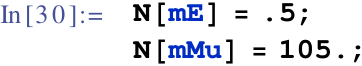}
\end{flushleft}

\tutorialStep Generate random numerical momenta for external particles (such that \lstinline!p1 + p2 = p3 + p4!):

\begin{flushleft}
    \includegraphics{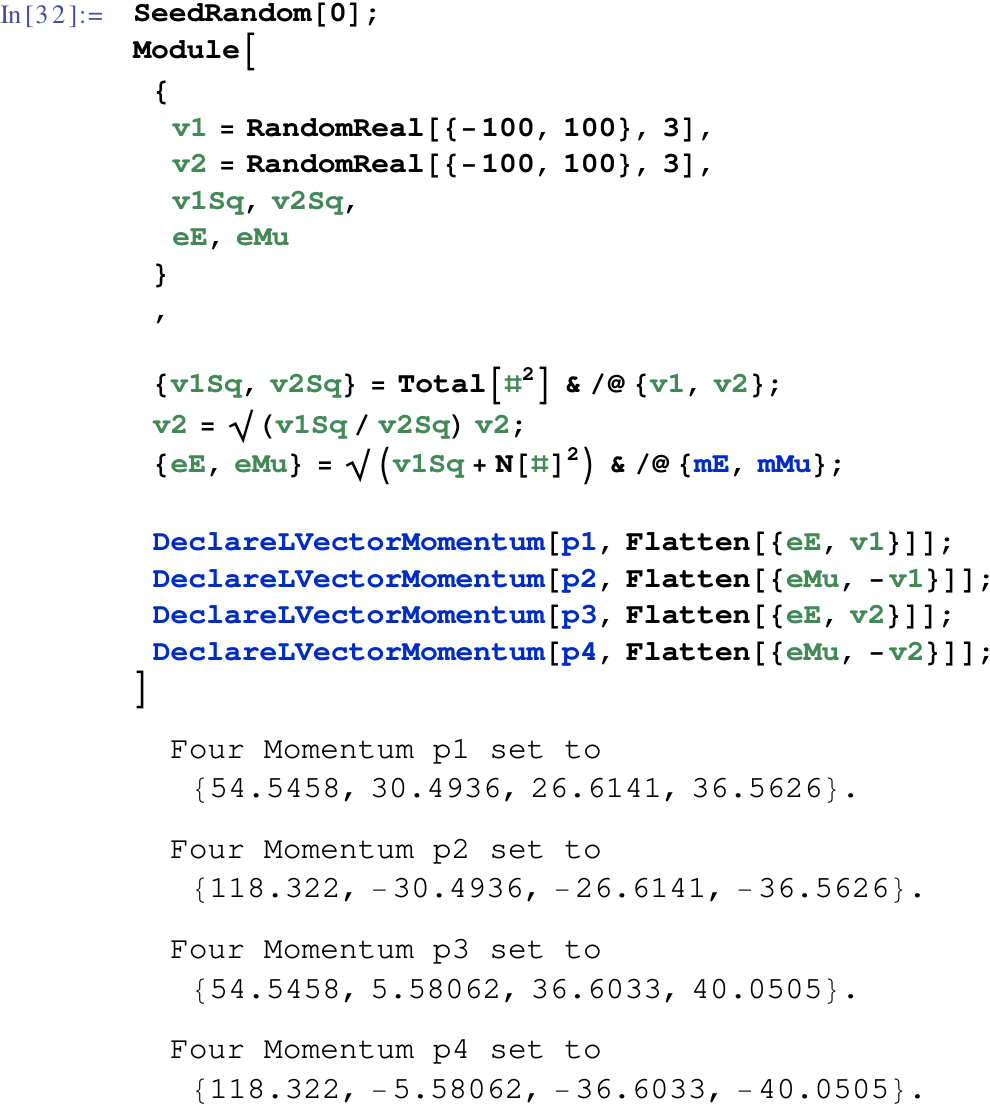}
\end{flushleft}

\tutorialStep Assign default numerical values to reference vectors for {``}final{''} particles:

\begin{flushleft}
    \includegraphics{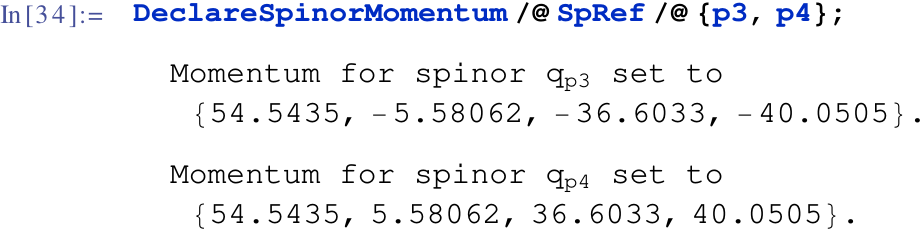}
\end{flushleft}

\tutorialStep Assign numerical values to associated vectors:

\begin{flushleft}
    \includegraphics{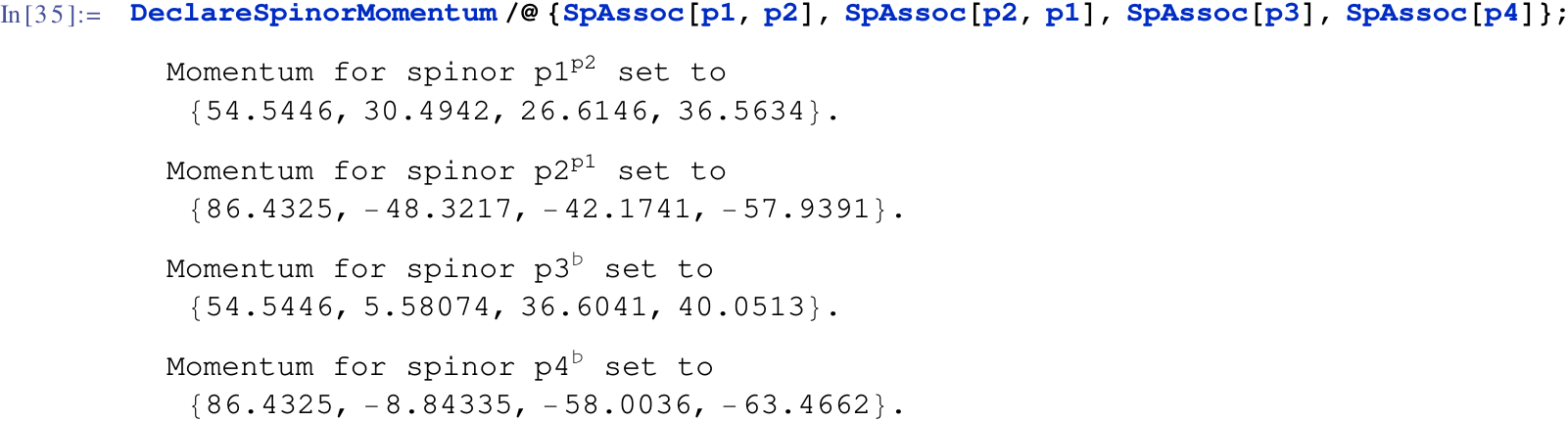}
\end{flushleft}

\subsubsection{Comparison with standard amplitude}

\tutorialStep Calculate electron muon scattering amplitude using ordinary Feynman diagrams (there is just one diagram: with off-shell photon \(\text{\textit{$t$}}\)-channel
exchange): 

\begin{flushleft}
    \includegraphics{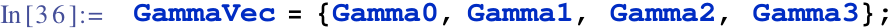}
\end{flushleft}

\begin{flushleft}
    \includegraphics{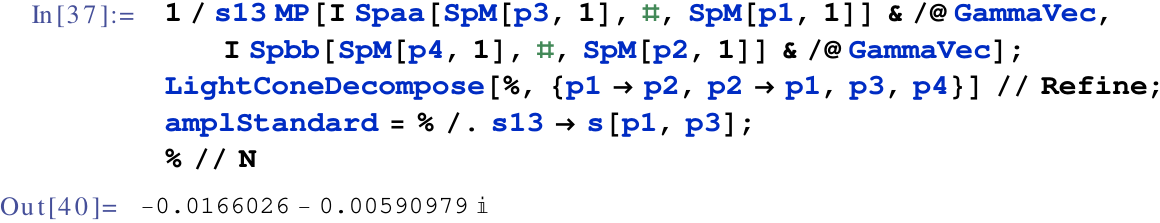}
\end{flushleft}

\tutorialStep Compute numerical value of amplitude calculated using on-shell recursion and compare it with amplitude calculated using Feynman diagrams:

\begin{flushleft}
    \includegraphics{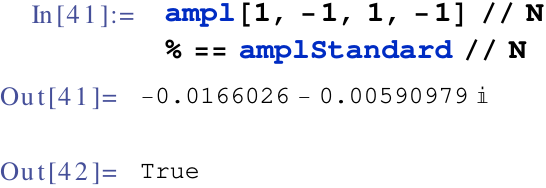}
\end{flushleft}

\subsubsection{Reference vectors independence}

\begin{table}
    \includegraphics{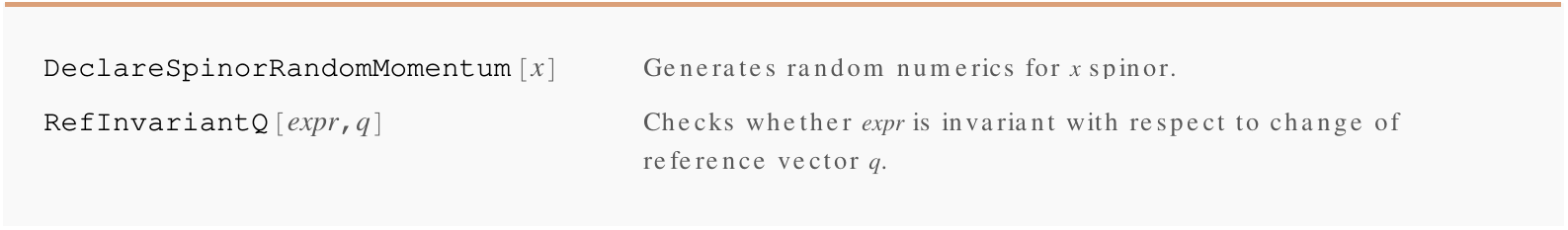}
    \caption{Functions used to test numerical invariance of amplitude.}
\end{table}

Independence of the final results on the choice of reference vectors can be used as additional test of the correctness of the final result:

\tutorialStep Assign random numerical values to reference vectors for exchanged photon:

\begin{flushleft}
    \includegraphics{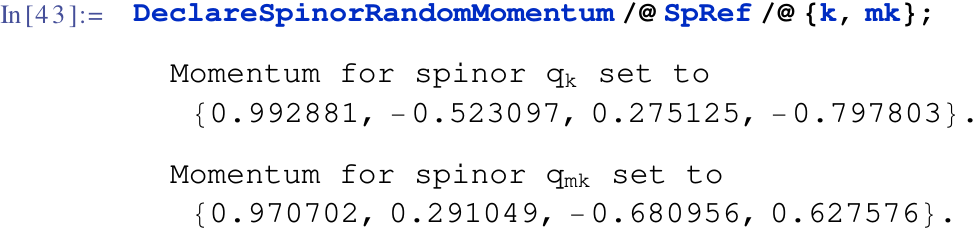}
\end{flushleft}

\tutorialStep Set accuracy of numerical invariance tests:

\begin{flushleft}
    \includegraphics{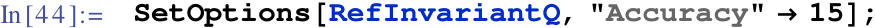}
\end{flushleft}

\tutorialStep Test invariance of diagrams with respect to change of reference vectors of on-shell photon:

\begin{flushleft}
    \includegraphics{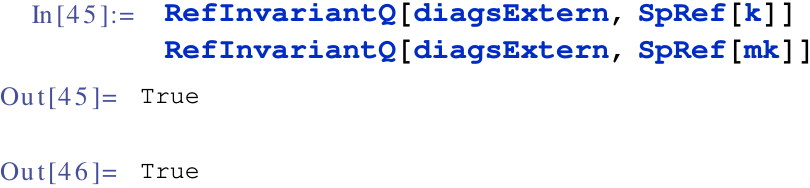}
\end{flushleft}

\tutorialStep Test invariance of amplitude with respect to change of reference vectors of external final particles (since different reference vectors
correspond to different, distinguishable states, amplitude is not invariant). Invariance testing function needs to take into account all occurrences
of reference vector, so we make all occurrences explicit.

\begin{flushleft}
    \includegraphics{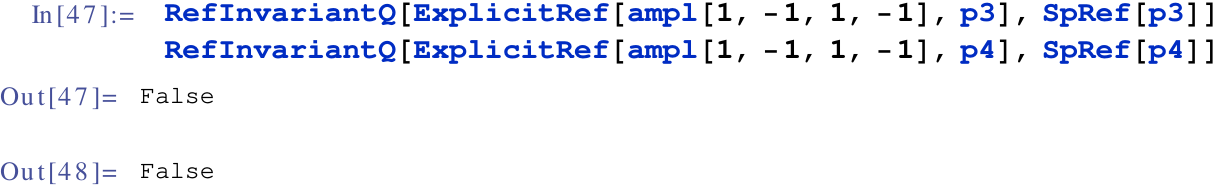}
\end{flushleft}

\tutorialStep Calculate amplitudes with changed spin projections of final particles:

\begin{flushleft}
    \includegraphics{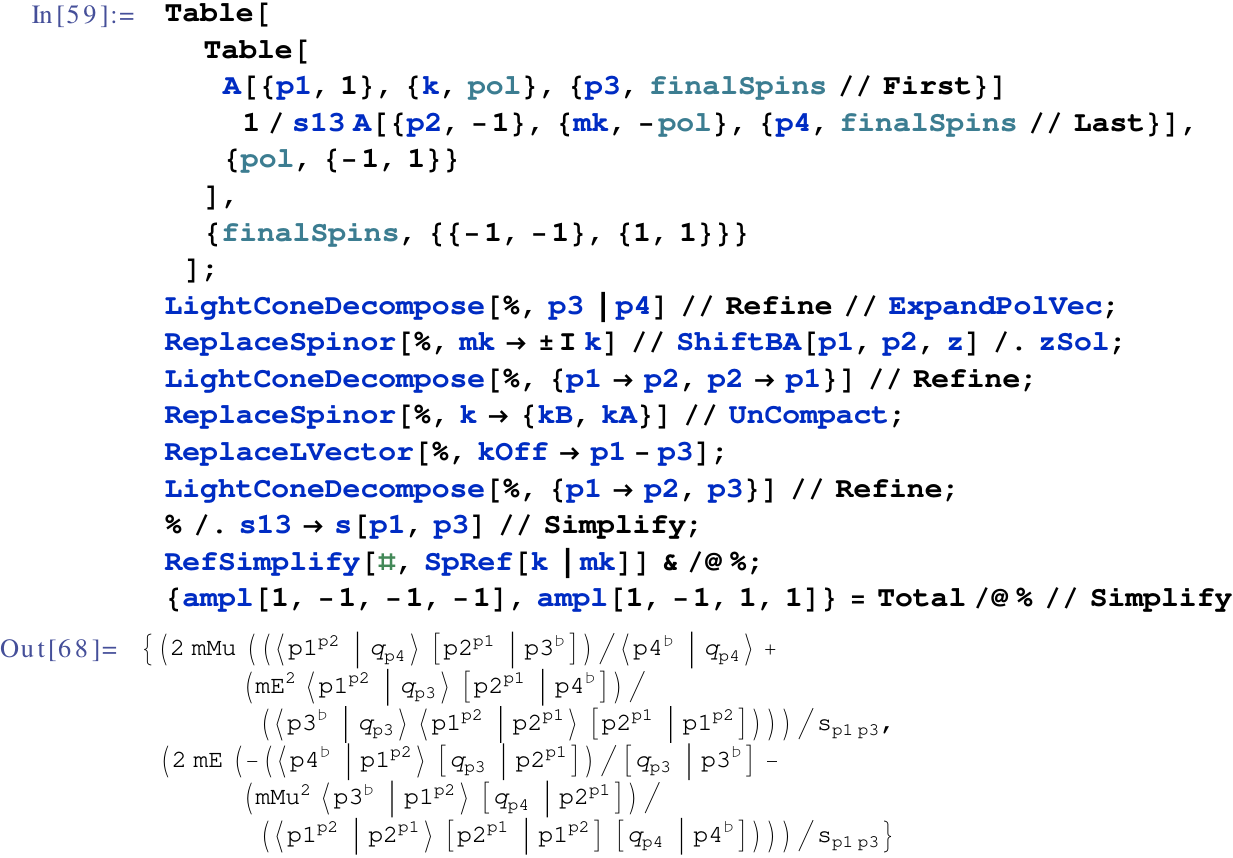}
\end{flushleft}

\tutorialStep Square of absolute value of amplitude summed over spin projections of given particle should be independent of reference vector related
to this particle:

\begin{flushleft}
    \includegraphics{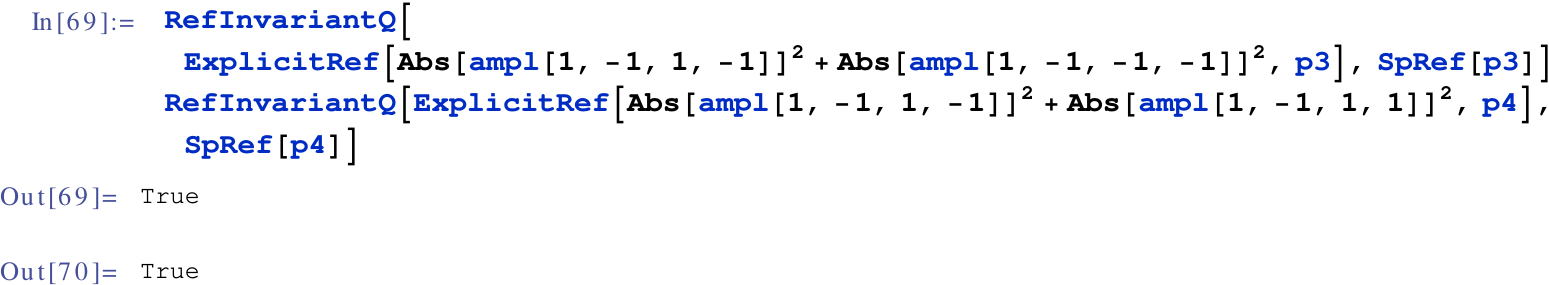}
\end{flushleft}

\section{Conclusions}
\label{sec:conclusions}

We have presented Mathematica package \texttt{SpinorsExtras} designed
for manipulating objects used in quantum field theory calculations in
the spinor helicity formalism.  SpinorsExtras is an extension of
already existing \texttt{S@M} package, enriching it with tools for
symbolic calculation of scattering amplitudes involving massive
particles.  Package was designed for computations using on-shell
recursion technique and implements complex shifts for massive spinors,
but is also suitable for calculations using other techniques, including
standard Feynman diagram computations. Thus, presented package can be also used
for comparison of results obtained with different methods. Like in the
\texttt{S@M} package, symbolic expressions calculated using
\texttt{SpinorsExtras} can be also evaluated numerically. Presented
package provides also utilities for testing gauge invariance and for
simplifying expressions with respect to change of reference vectors.

\subsection*{Acknowledgments}
\label{sec:acknowledgments}

Author would like to thank Janusz Rosiek for useful discussions and
comments on the manuscript.  This work has been supported by Polish
National Science Centre under the research grants
DEC-2011/01/M/ST2/02466 and DEC-2012/05/B/ST2/02597.

\newpage
\bibliography{SpinorsExtras}
\bibliographystyle{utphys}

\end{document}